\documentclass[preprint,nofootinbib,amsmath,amssymb,aps,floatfix]{revtex4-1}

\pdfoutput = 1

\usepackage{graphicx}
\usepackage{dcolumn} 
\usepackage{multirow}
\usepackage{bm}
\usepackage{color}
\usepackage[table]{xcolor}
\usepackage{subfig}
\usepackage[utf8]{inputenc}
\usepackage[toc,page]{appendix}
\usepackage{xspace}

\definecolor{darkblue}{rgb}{0.0,0,0.5}
\definecolor{darkred}{rgb}{0.7,0,0.0}
\definecolor{medmagenta}{rgb}{0.3,0,0.7}
\definecolor{darkmagenta}{rgb}{0.4,0,0.4}
\definecolor{lightgray}{gray}{0.87}
\usepackage[unicode=true, bookmarks=false, linkcolor = darkblue, citecolor = medmagenta, urlcolor=darkmagenta,breaklinks=false, colorlinks=true, hyperfootnotes=true]{hyperref}
\DeclareUnicodeCharacter{FF0C}{,}

\usepackage{epsfig,amssymb,amsmath,psfrag,epstopdf,color}
\usepackage{ulem}
\usepackage{amsthm}

\def\sqs{\ensuremath{\protect\sqrt{s}}\xspace}
\def\invfb{\ensuremath{\mbox{\,fb}^{-1}}\xspace}
\def\pt{\mbox{$p_{\mathrm{ T}}$}\xspace}
\def\doveru{\mbox{$d/u$}\xspace}
\def\wz {\mbox{$W^{\pm}$ and $Z$}\xspace}
\def\doverubar {\mbox{${\bar{d}}/{\bar{u}}$}\xspace}
\def\ssoverud {\mbox{$\frac{s+\bar{s}}{\bar{u}+\bar{d}}$}\xspace}

\newcommand{\tev}{\ensuremath{\mathrm{\,Te\kern -0.1em V}}\xspace}
\newcommand{\gevc}{\ensuremath{{\mathrm{\,Ge\kern -0.1em V\!/}c}}\xspace}
\newcommand{\gevcc}{\ensuremath{{\mathrm{\,Ge\kern -0.1em V\!/}c^2}}\xspace}
\begin{document}

\preprint{MSUHEP-20-015}

\title{Impact of LHCb 13 TeV $\bold{W}$ and $\bold{Z}$ pseudo-data on the parton distribution functions}

\author{Qilin Deng}
\affiliation{
Institute of Particle Physics, Central China Normal University,\\
 Wuhan, Hubei 430079 China }
 
\author{Sayipjamal Dulat}
\email{sdulat@hotmail.com}
\affiliation{
School of Physics Science and Technology, Xinjiang University,
 Urumqi, Xinjiang 830046 China }

\author{Qundong Han}
\affiliation{
Institute of Particle Physics, Central China Normal University,\\
 Wuhan, Hubei 430079 China }
 
\author{Tie-Jiun Hou}
\email{houtiejiun@mail.neu.edu.cn}
\affiliation{
Department of Physics, College of Sciences, Northeastern University,
 Shenyang, Liaoning 110819 China }

\author{Hang Yin}
\email{yinh@ccnu.edu.cn}
\affiliation{
Institute of Particle Physics, Central China Normal University,\\
 Wuhan, Hubei 430079 China }

\author{ C.--P. Yuan}
\email{yuan@pa.msu.edu}
\affiliation{
Department of Physics and Astronomy, Michigan State University,
 East Lansing, MI 48824 U.S.A. }

\date{September 8, 2020}
%\date{\today}

%%%%%%%%%%%%%%%%%%%%%%%%%%%%%%%%%%%%%%%%%%%%%%%%

\begin{abstract}
We study the potential of the LHCb 13\tev single \wz boson pseudo-data 
for constraining the parton distribution functions (PDFs) of the proton. As an example, we demonstrate the sensitivity of the LHCb 13\tev data, 
collected with integrated luminosities of 5\invfb and 300\invfb, to reducing the PDF uncertainty bands of the CT14HERA2 PDFs, 
using the error PDF updating package {\sc ePump}. The sensitivities of various experimental observables are compared. 
Generally, sizable reductions in PDF uncertainties can be observed in the 300\invfb data sample, particularly in the small-$x$ region. 
The double-differential cross section measurement of $Z$ boson \pt and rapidity can greatly reduce the uncertainty bands of $u$ and $d$ 
quarks in almost the whole $x$ range, as compared to various single observable measurements. 
\end{abstract}

\pacs{12.15.Ji, 12.15.Lk, 13.85.Qk}
\keywords{PDF; W/Z boson production;}
%%%%%%%%%%%%%%%%%%%%%%%%%%%%%%%%%%%%%%%%%%%%%%%%

\maketitle

\newpage
\tableofcontents
\newpage

%%%%%%%%%%%%%%%%%%%%%%%%%%%%%%%%%%%%%%%%%%%%%%%%
\section{Introduction} \label{sec:Introduction}
In hadron colliders, most 
physics analyses rely heavily on the understanding of the 
parton picture of hadronic beam particles, including the precision measurements of the 
Standard Model (SM) parameters~\cite{Aaltonen:2012bp,Abazov:2012bv,Aaboud:2017svj}, and new physics searches. 
The parton picture follows the factorization theorem of quantum chromodynamics (QCD).
The parton distribution functions (PDFs) are nonperturbative, and therefore cannot be calculated.
They are functions of the Bjorken-$x$ values ($x$, momentum fraction) of partons at a momentum transfer scale ($Q$),
which are determined phenomenologically by a global analysis of experimental data from a 
wide range of physics processes, such as Deep Inelastic Scattering (DIS), Drell-Yan (DY), inclusive jets, and top quark pair production processes. 
The PDF dependencies on $Q$ are determined by the renormalization-group based evolution equations, {\it e.g.}~DGLAP equation~\cite{Gribov:1972ri,Dokshitzer:1977sg,Altarelli:1977zs}.

Precision measurements of the single $W^\pm$ and $Z$ gauge boson production cross section\footnote{Throughout this paper, $Z$ includes both the $Z$ boson and the virtual photon contributions.} at the CERN Large Hadron Collider (LHC) provide
important tests of the 
QCD and the electroweak (EW) sectors of the SM. 
Theoretical predictions for these cross sections are available up to next-to-next-to-leading order 
(NNLO) in perturbative QCD~\cite{Rijken:1994sh,Hamberg:1990np,Harlander:2002wh,vanNeerven:1991gh,Anastasiou:2003ds}, 
where one of the dominated systematic uncertainties comes from the PDFs.
The CT14 PDFs~\cite{Dulat:2015mca} are the first CTEQ-TEA PDFs that include 
published results from the ATLAS, CMS, and LHCb collaborations at 7\tev,
including the \wz gauge boson production
cross sections and the lepton charge asymmetry measurements from the ATLAS Collaboration~\cite{Aad:2011dm}, 
the lepton charge asymmetry in the electron~\cite{Chatrchyan:2012xt} 
and muon channels~\cite{Chatrchyan:2013mza} from the CMS Collaboration, 
the lepton charge asymmetry in the decay of $W^\pm$-bosons to an electron or a muon, 
and the $Z$ boson rapidity distribution from 
the LHCb Collaboration~\cite{Aaij:2012vn}. 
The ATLAS and CMS measurements primarily impose constraints on the light quark and antiquark PDFs at
$x \gtrsim 0.01$. 
As studied in Refs.~\cite{Alekhin:2013nda,Gao:2017yyd}, 
the LHCb 7\tev and 8\tev \wz boson measurements, 
though with larger statistical uncertainties as compared to the corresponding results from the ATLAS and CMS experiments,
could also impose significant constraints on $u$ and $d$ PDFs.

\begin{figure}[!htbp]
\centering
\includegraphics[width=0.45\textwidth]{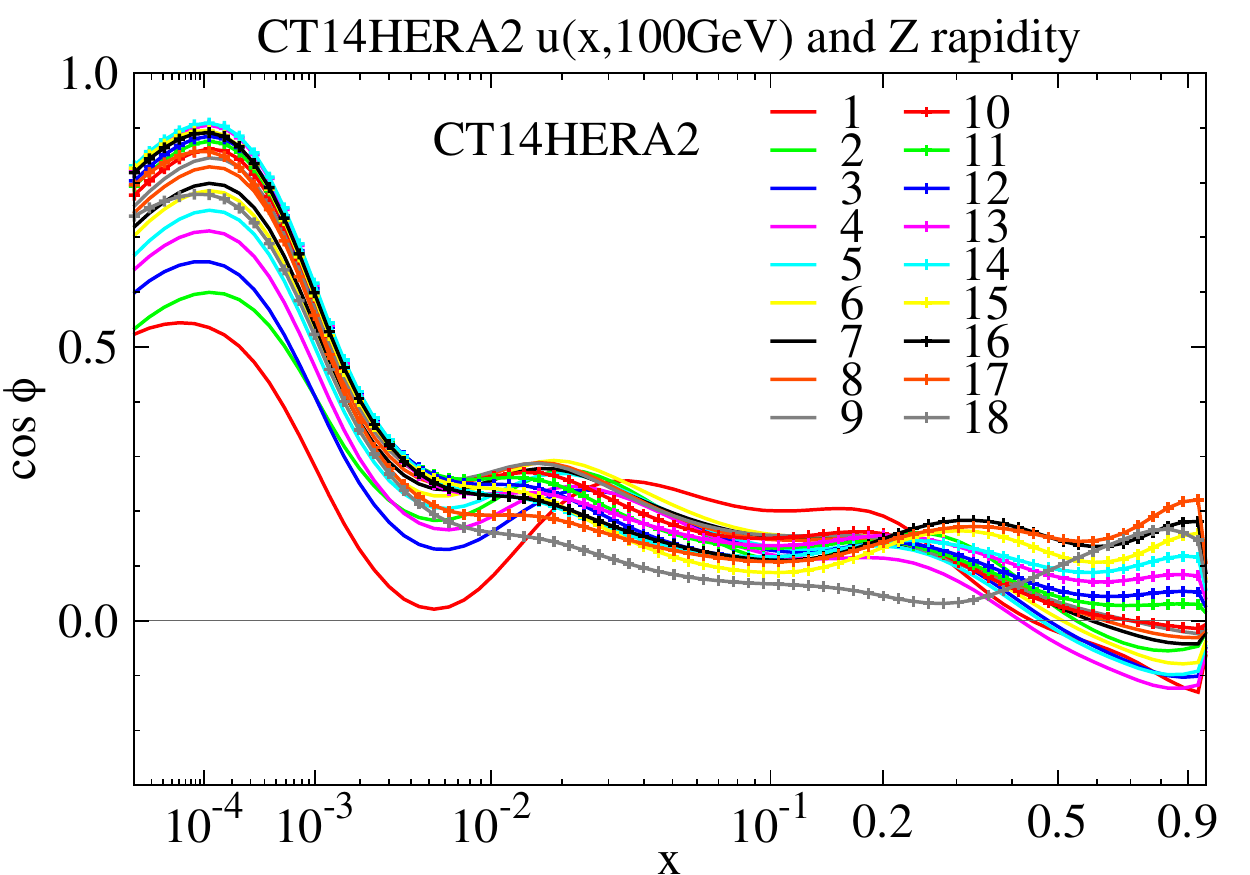}
\includegraphics[width=0.45\textwidth]{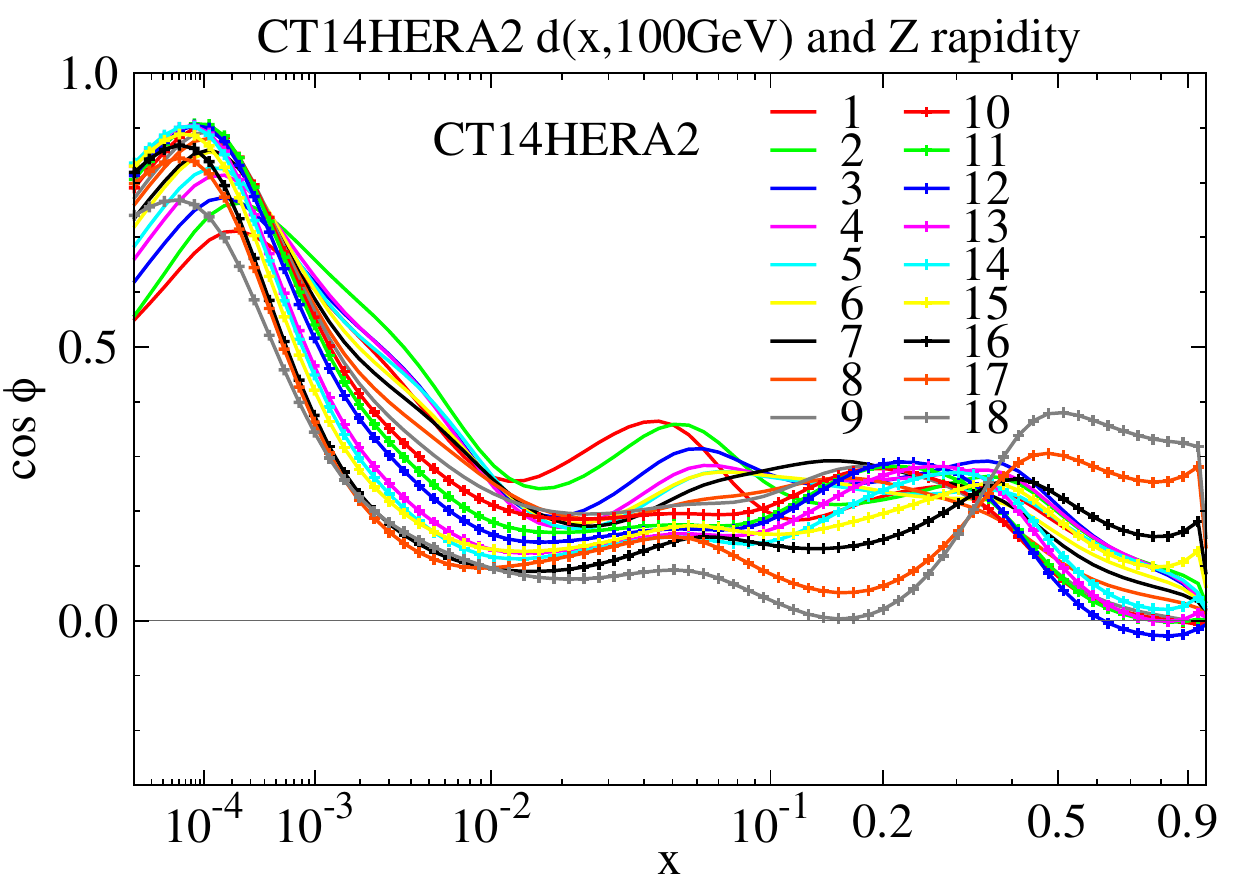}
\caption{Correlation $\cos \phi$~\cite{Nadolsky:2008zw} between (left) $u(x,Q)$ PDF or (right) $d(x,Q)$ PDF, and 
the DY differential cross-section in 18 bins of $Z$ boson rapidity, as predicted by {\sc ResBos} 
with CT14HERA2 PDFs, at $Q = 100\,{\rm GeV}$.
The same binning scheme as the LHCb paper~\cite{Aaij:2012vn} is used.
\label{fig:z_x_corr}}
\end{figure}

In the past few decades, a large number of experimental results have been used in the PDF global analysis, 
but we still have limited knowledge of the PDFs in very small- and very large-$x$ ranges.
In the single \wz boson production, the $x$ value of interacting partons ($x_1$ and $x_2$) are correlated with the boson
production, through its rapidity ($y$), as $y=\frac{1}{2}\ln\frac{x_1}{x_2}$. Therefore, the single \wz data 
in the forward detector region are valuable in the PDF global analysis, as events with larger boson rapidity 
are produced by partons with small or large $x$. 
Correlations between the predicted LHCb 13\tev $Z$ boson production cross section and the $u$ and $d$ quark PDFs as a function of Bjorken-$x$ are shown in Fig.~\ref{fig:z_x_corr}.
As shown in the figure, the LHCb 13\tev data is expected to have strong correlations with $u$ and $d$ quarks in the small-$x$ region, 
indicating the LHCb 13\tev \wz data can be used to constrain the corresponding PDFs.

The LHCb detector~\cite{Alves:2008zz,Aaij:2014jba} is a single-arm forward spectrometer designed for the study of particles containing $b$ or $c$ quarks,
covering the pseudorapidity range $2<\eta<5$. 
With a high-performance tracking system and a muon sub-detector, 
the LHCb data can also be extended to precision EW measurements. 
Using $pp$ collision data collected at ${\sqs=7\tev}$,
the LHCb has measured the single \wz boson production cross sections using both muon and electron
channel~\cite{Aaij:2012vn,Aaij:2012mda,Aaij:2014wba}, and the same measurements had been
performed using ${\sqs=8\tev}$ data~\cite{Aaij:2015zlq,Aaij:2015vua,Aaij:2016qqz}.
These results have been used to constrain the PDFs~\cite{Harland-Lang:2014zoa,Ball:2014uwa,Dulat:2015mca},
bring valuable information to the PDF analysis.
The ${\sqs=13\tev}$ $pp$ collision data sample has been collected with a
larger center-of-mass energy than previous publications, 
with more \wz boson events boosted to the forward region, and therefore allows access to
even smaller (or larger) values of $x$ than the previous 7 and 8\tev results.

The LHCb detector runs at reduced luminosity compared to the ATLAS and CMS detectors~\cite{Aaij:2014ida},
because the detector occupancy is extremely high in the forward region.
In the LHC Run 2 period (2015--2018), the LHCb detector collected more than 5\invfb $pp$ collision data~\cite{Aaij:2019kcg} at ${\sqs=13\tev}$.
An upgraded detector~\cite{lhcb:upgrade1} is foreseen to allow LHCb detector operation at a 
luminosity of $2\times 10^{33}$ $\text{cm}^{-2}\text{s}^{-1}$ in the LHC Run 3 period (2022-2024),
which is an instantaneous luminosity five times higher than before. 
By the end of the LHC Run 4 period (2026-2029), the LHCb detector is expect to collect approximately 50\invfb $pp$ collision data~\cite{lhcb:upgrade2}.
There is also a plan for a future LHCb Upgrade-II phase (planned for 2031 data taking)~\cite{lhcb:u2phys}, to guarantee
the LHCb detector could run at an even higher luminosity ({$2\times 10^{34}$ $\text{cm}^{-2}\text{s}^{-1}$})~\cite{lhcb:highL}.
After the LHCb detector Upgrade-II, by the end of the LHC operation, it is planned that the 
LHCb detector should have collected a data sample corresponding to a minimum 300\invfb~\cite{lhcb:upgrade2}. 
Therefore, in this article, the pseudo-data samples used for the 
physics projections are set to either 5\invfb or 300\invfb. 

The article is organized as follows. In Sec.~\ref{sec:epump}, 
we discuss the error PDF updating package {\sc ePump} and pseudo-data samples used in the analysis.
In Sec.~\ref{sec:LHCbimpact}, we study the impacts of the LHCb 13\tev single \wz boson pseudo-data on the CT14HERA2 PDFs~\cite{Dulat:2015mca,Hou:2016nqm}.
In Sec.~\ref{sec:tolerance}, we discuss the choice of tolerance criteria in the {\sc ePump} update.
Our conclusion is given in Sec.~\ref{sec:summary}.

%%%%%%%%%%%%%%%%%%%%%%%%%%%%%%%%%%%%%%%%%%%%%%%%
\section{ Updating Error PDFs}\label{sec:epump}
 
\begin{figure}[!htbp]
\centering
\includegraphics[width=0.45\textwidth]{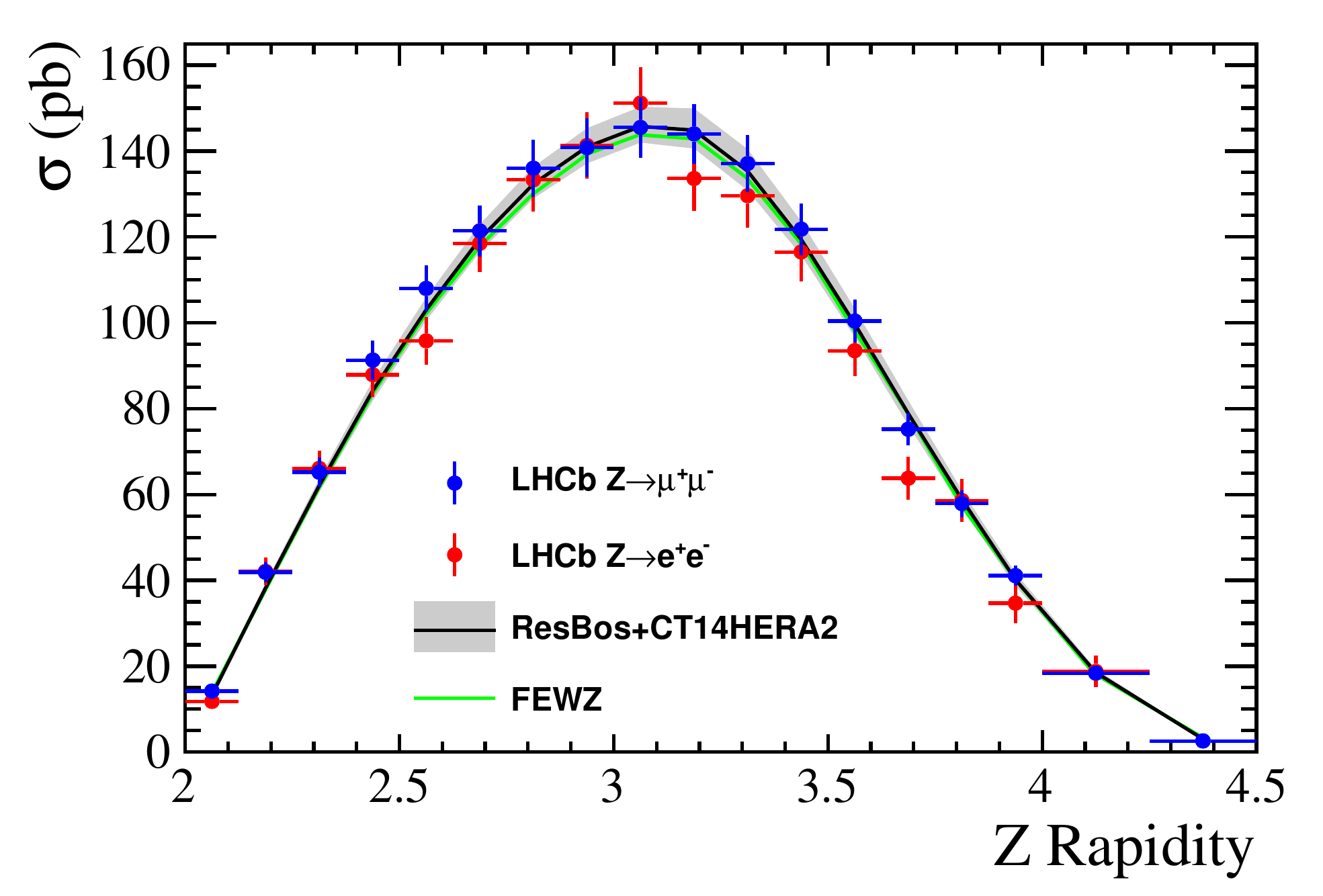}
\includegraphics[width=0.45\textwidth]{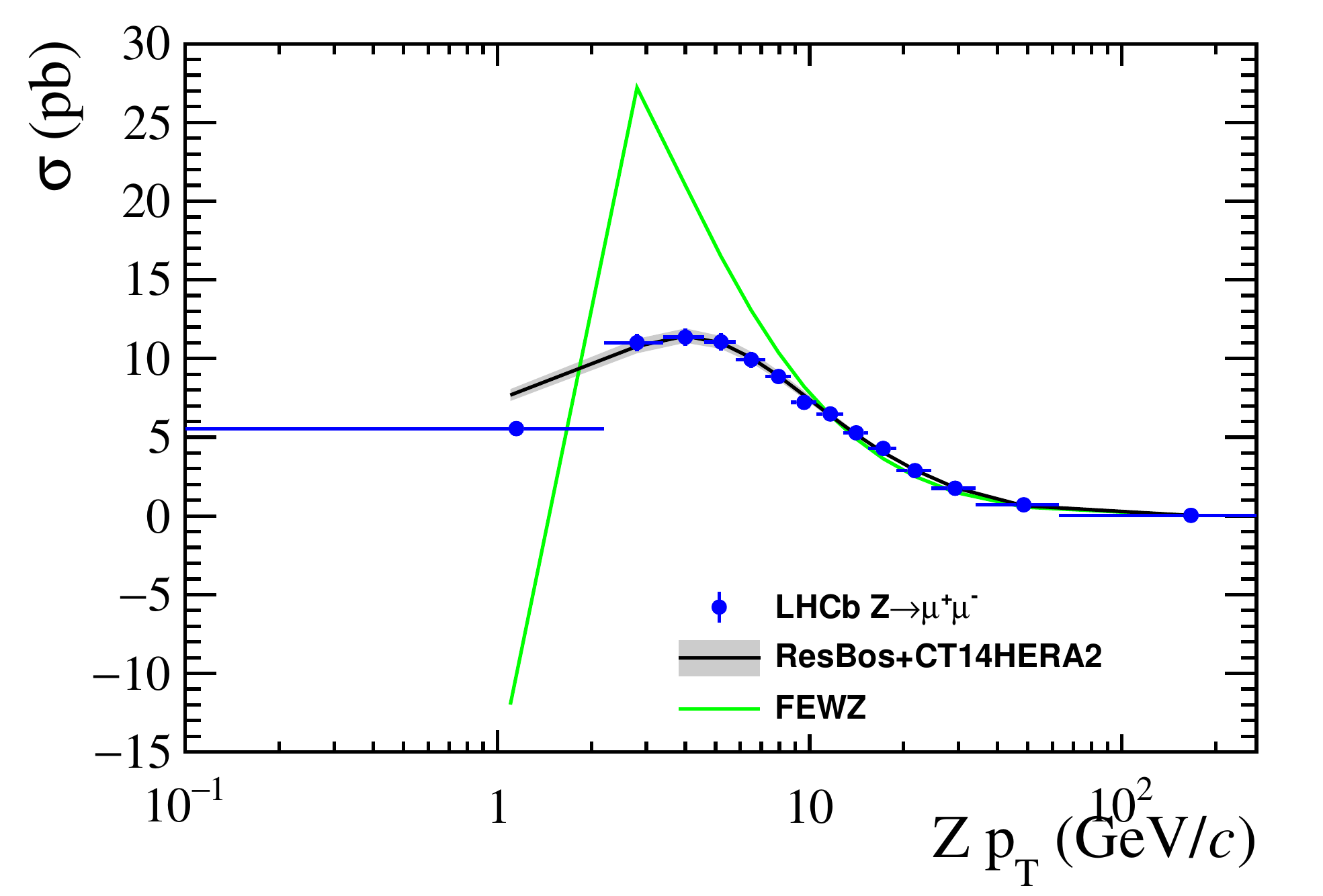}
\includegraphics[width=0.45\textwidth]{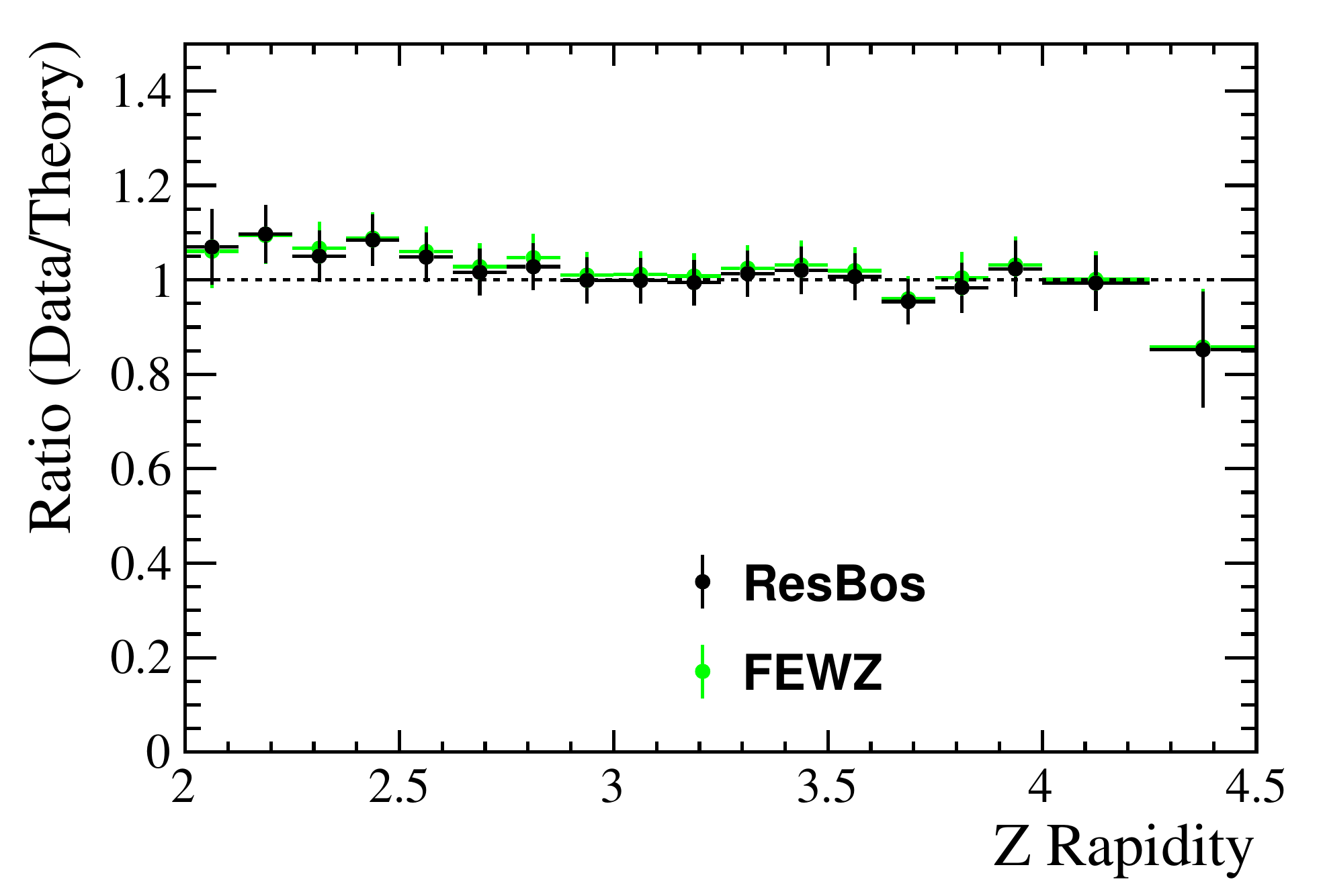}
\includegraphics[width=0.45\textwidth]{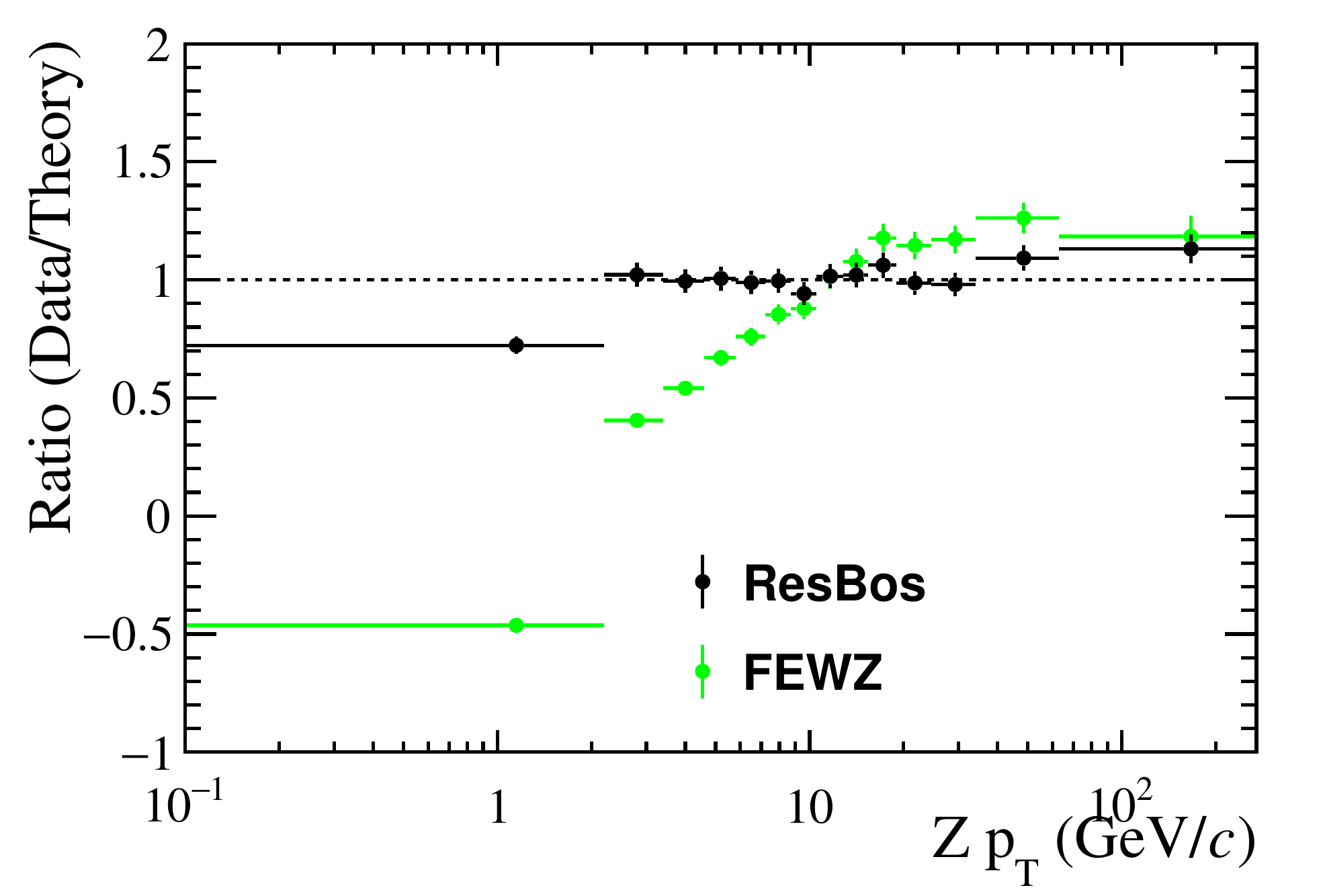}
\caption{
Comparison of DY differential cross section as a function of the $Z$ boson rapidity (top-left) and \pt (top-right), between theory ({\sc ResBos}, and {\sc FEWZ}) and the LHCb 13\tev data~\cite{Aaij:2016mgv}. The blue points represent the LHCb results in muon channel, and the red points the electron channel, while the black (green) line represents the {\sc ResBos} ({\sc FEWZ}) prediction, the grey band represents the PDF uncertainty at the 68\% confidence level (CL) estimated using the CT14HERA2 PDF set. The ratio of data to theoretical prediction are shown as a function of $Z$ boson rapidity (bottom-left) and \pt (bottom-right).}
\label{fig:lhcb_comp}
\end{figure}

Recently, the CTEQ-TEA global analysis group has released a tool named {\sc ePump} (Error PDF Updating Method Package)~\cite{Schmidt:2018hvu,Hou:2019gfw}, 
which can be used to explore the impact of new data on existing PDFs, without perform a full global analysis. 
The {\sc ePump} method has been demonstrated in Ref.~\cite{Willis:2018yln}. 
In the {\sc ePump} update, two inputs are needed: a measured result from experiment and the corresponding theoretical predictions of the complete set of error PDFs. 
Thus far, the LHCb Collaboration has only published the single $Z$ boson production cross section~\cite{Aaij:2016mgv} at 13\tev,
using a small fraction of its Run 2 data sample (2015 data, with an integrated luminosity of 0.3\invfb).
The comparisons between the LHCb 13\tev data and {\sc ResBos} and {\sc FEWZ}~\cite{Li:2012wna} predictions are shown in Fig.~\ref{fig:lhcb_comp},
with good agreement between data and {\sc ResBos} prediction.
{\sc FEWZ} can provides fixed order
calculation at NNLO accuracy in QCD. In this paper, the {\sc FEWZ} prediction is calculated with the same parameter setting 
as the previous LHCb publication~\cite{Aaij:2016mgv}, using the CT14NNLO PDF set.
However, there is no publication for the single $W^{\pm}$ boson production using the 13\tev LHCb data. 
Therefore, we shall use pseudo-data in this analysis to emulate the impact of the upcoming LHCb 13\tev data on the PDFs.

The Monte Carlo events generated with the {\sc ResBos} generator~\cite{Balazs:1997xd}, 
using the MMHT14~\cite{Harland-Lang:2014zoa} and CT14HERA2~\cite{Dulat:2015mca,Hou:2016nqm} PDFs, are taken as
the pseudo-data and theoretical predictions, respectively, in this work. 
The theoretical predictions in this work are computed using the {\sc ResBos}~\cite{Balazs:1997xd} package at approximate 
NNLO plus next-to-next-to-leading-logarithm (NNLL) in QCD interactions, in which the canonical scales are used~\cite{Landry:2002ix,Su:2014wpa}.
For example, both the renormalization and factorization scales are set to be the invariant mass of the lepton pair in the Drell-Yan (DY) events. 

To emulate the LHCb detector acceptance, the charged leptons ($\ell^{\pm}$, electrons or muons) are required to have transverse momentum ($p_T$) 
greater than $20\gevc$ and pseudorapidity ($\eta$) in the range of
$2.0<\eta<4.5$. In the case of the $Z$ boson events, the invariant mass of the dilepton pair is required to be in the range from 60\gevcc to 120\gevcc.

In the {\sc ePump} study, for the $Z\rightarrow \ell^+\ell^-$ pseudo-data input, 
the statistical uncertainties are scaled to the 5\invfb and 300\invfb data samples, separately, 
by extrapolating the total uncertainty of the LHCb 13\tev publication~\cite{Aaij:2016mgv} to the pseudo-data sample.
In the extrapolation, an assumption is made that the ratio of statistical 
uncertainty to systematical uncertainty will remain the same in all data samples. 
Similarly, for the $W^{\pm}\rightarrow \ell^{\pm} \nu$ pseudo-data sample, 
uncertainties are estimated using the LHCb 8\tev publication~\cite{Aaij:2015zlq,Aaij:2016qqz}.

%%%%%%%%%%%%%%%%%%%%%%%%%%%%%%%%%%%%%%%%%%%%%%%%
\section{Impact of the LHCb 13 TeV $W^\pm $ and $Z$ pseudo-data on CT14HERA2 PDFs}
\label{sec:LHCbimpact}

In this section, we study the impact of the LHCb 13\tev single \wz boson pseudo-data 
on the CT14HERA2 PDFs, to demonstrate the LHCb 13\tev data sensitivity,
and to further investigate valuable observables for future measurements.

Since the pseudo-data sample generated with the MMHT14 PDFs is used to update the 
CT14HERA2 PDF sets, and there are differences between the central PDF 
set of MMHT14 and CT14HERA2, 
the central value of {\sc ePump} updated-PDFs varies from the CT14HERA2 central set.
In this article, we are interested in variations of PDF uncertainty, and thus we do not discuss variations of PDF central values hereafter. 

%%%%%%%%%%%%%%%%%%%%%%%%%%%%%%%%%%%%%%%%%%%%%%%%
\subsection{Update from LHCb 13 TeV $W^\pm$ pseudo-data} \label{subsec:w}
 
In the $W^{\pm}$ boson leptonic decay, there is a neutrino and a charged lepton in the final state, where the neutrino will escape from detector, 
only the charged lepton can be detected in a hadron collider experiment.
This feature makes a $W^{\pm}$ boson analysis complicated, since the irreducible background 
contribution is difficult to model.
On the other hand, the single $W^\pm$ production rate is one order of magnitude larger than that of 
the $Z$ boson at the LHCb. 
If we could model the background properly for the $W^{\pm}$ events, 
such a sample with large statistics could allow us to perform many precision measurements.
In this study, we used the charged lepton pseudorapidity distribution ($\eta$) as an observable in the {\sc ePump} update, 
with the same binning scheme as the previous LHCb publications~\cite{Aaij:2015zlq,Aaij:2016qqz}. 

\begin{figure}[!htbp]
\centering
\includegraphics[width=0.45\textwidth]{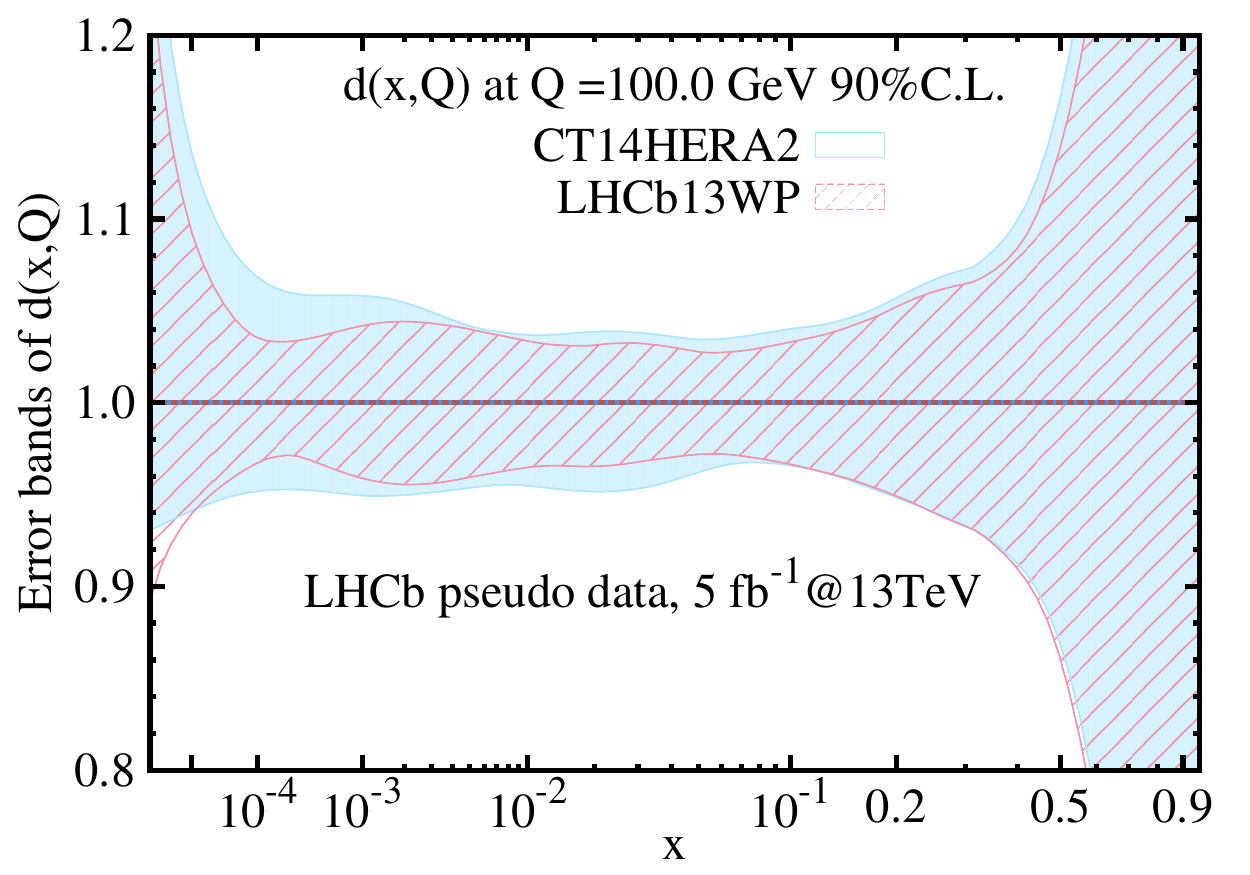}
\includegraphics[width=0.45\textwidth]{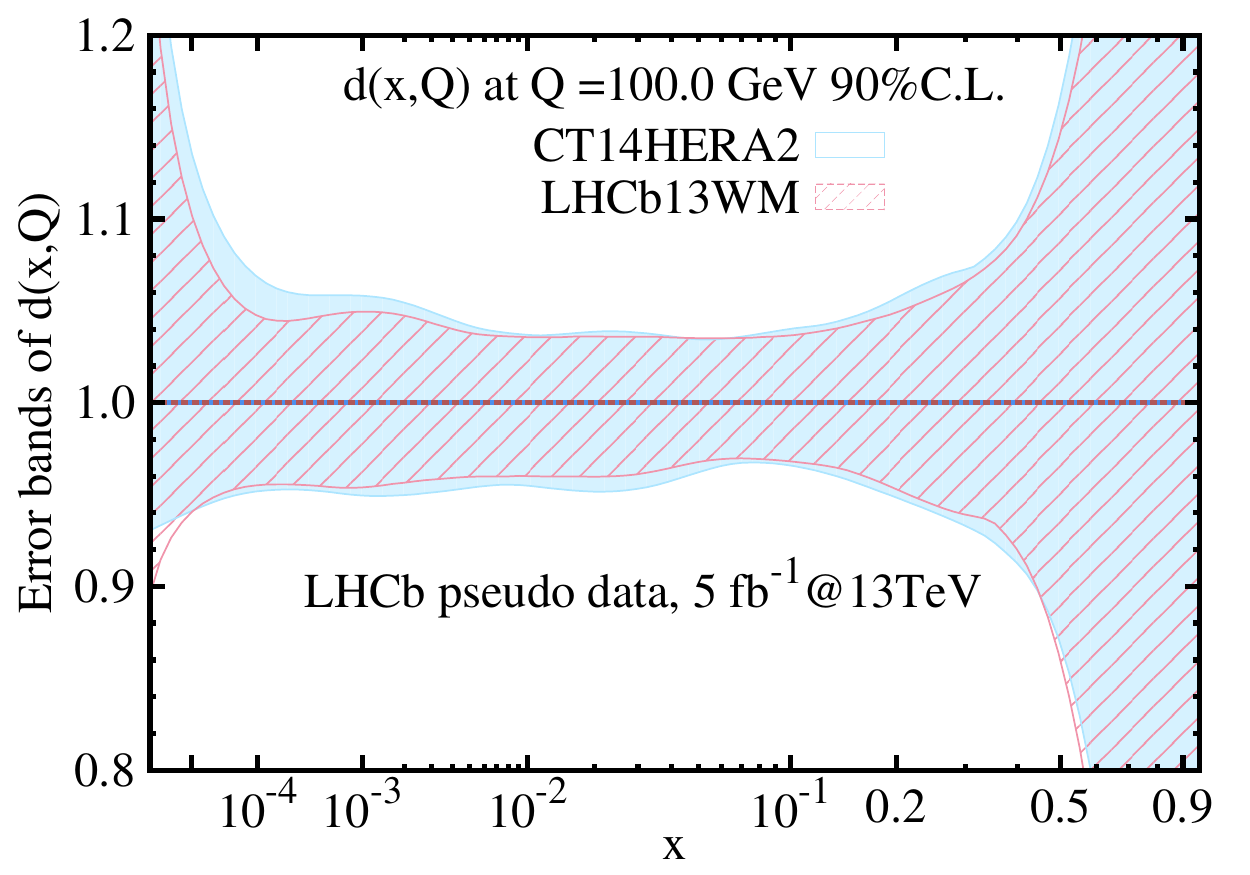}
\includegraphics[width=0.45\textwidth]{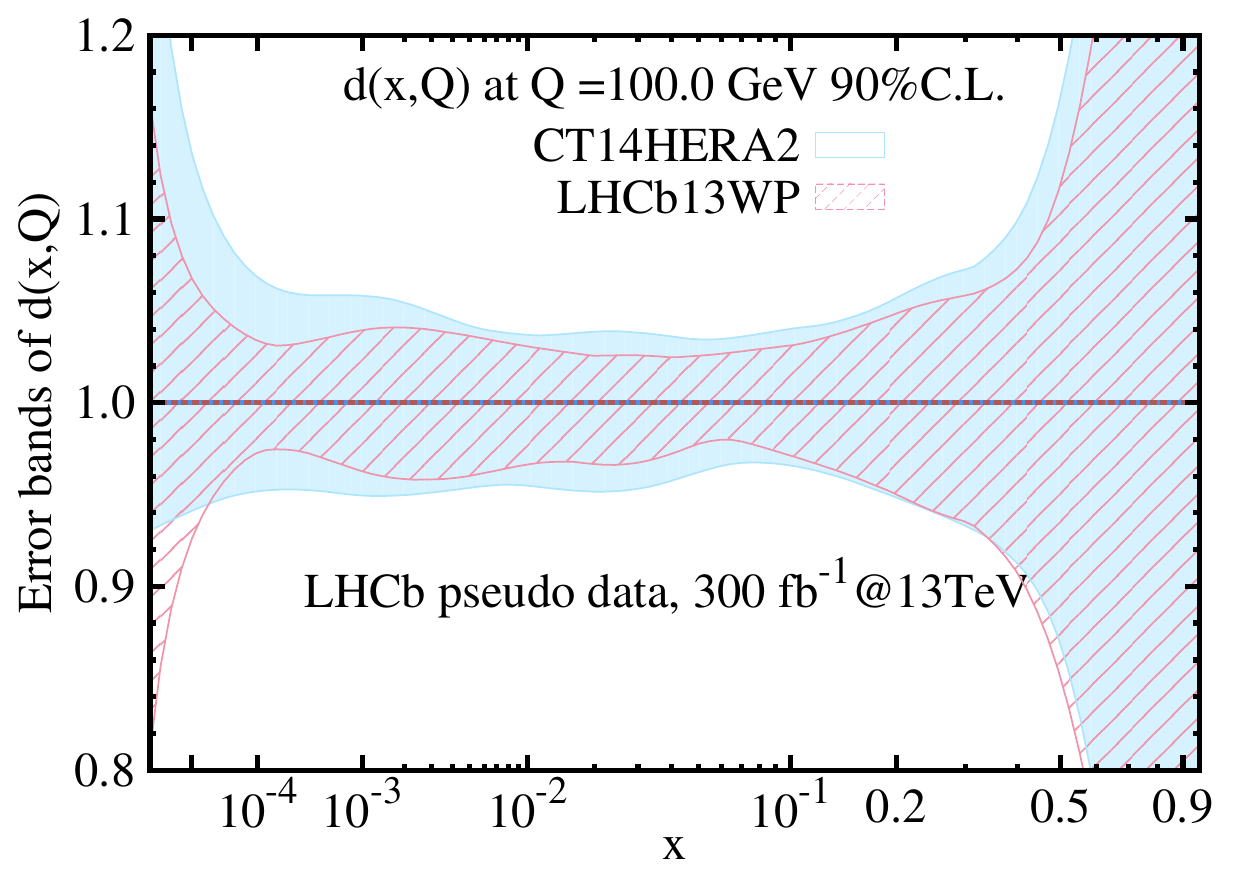}
\includegraphics[width=0.45\textwidth]{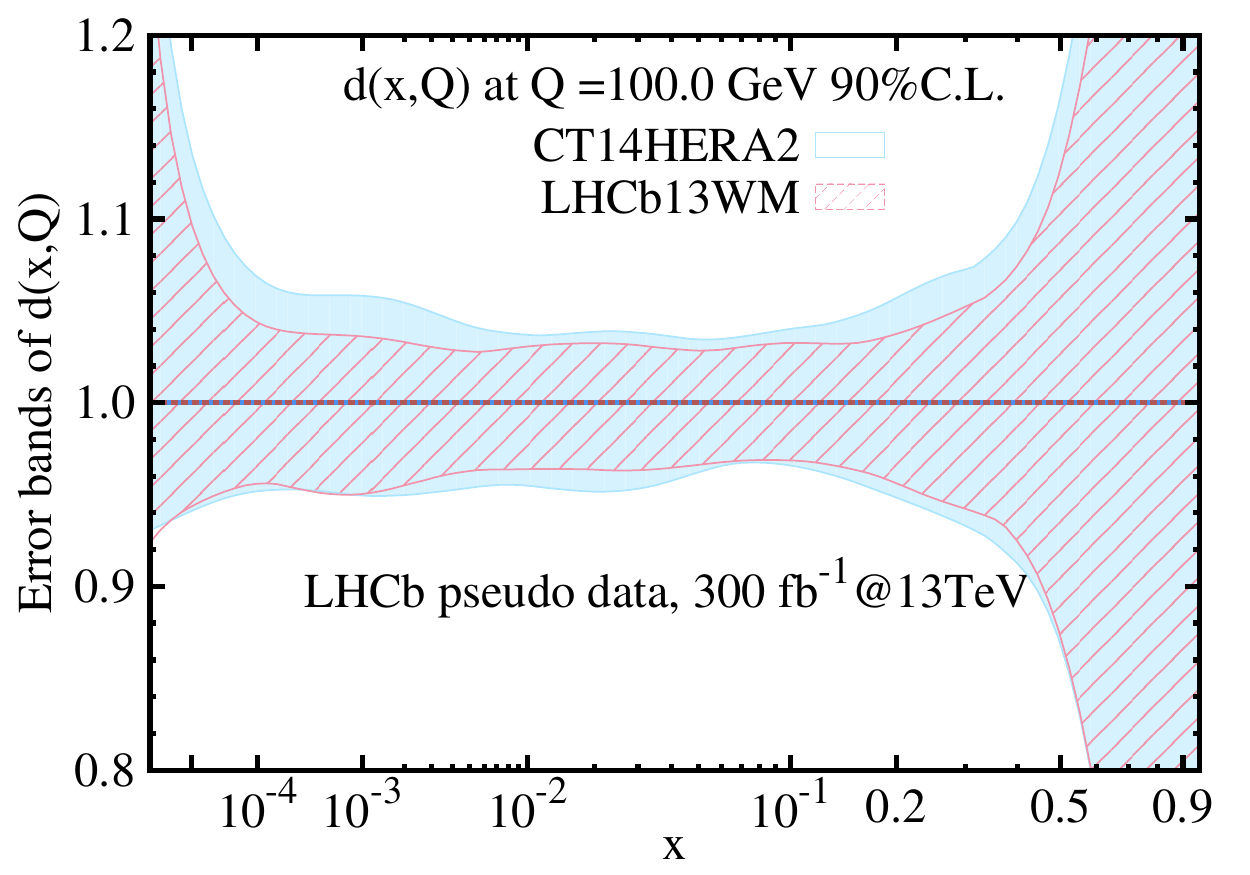}
\caption{PDF uncertainties associated with the $d$ quark, as a function of $x$, in the CT14HERA2 PDFs and {\sc ePump} updated new PDFs. 
The denominator is the central value of each PDF set. The blue (red) line represents the central value of the PDF ratio before (after) the {\sc ePump} update, the blue band represents the CT14HERA2 PDF uncertainty, the red shaded band represents the updated PDF uncertainty. The charged lepton pseudorapidity distributions of $W^{\pm}$ events are used as inputs for {\sc ePump} update. 
(top-left) The $d$ quark result using 5\invfb $W^+$ events;
(top-right) the $d$ quark result using 5\invfb $W^-$ events;
(bottom-left) the $d$ quark result using 300\invfb $W^+$ events,
and (bottom-right) the $d$ quark result using 300\invfb $W^-$ events.}
\label{fig:wpm_dquark}
\end{figure}

After the {\sc ePump} update, the updated quark PDF distribution
is compared with the default one of CT14HERA2.
The $d$ quark PDF distribution with its uncertainty is shown in Fig.~\ref{fig:wpm_dquark}, 
which indicates that the LHCb 13\tev $W^\pm$ boson data have a large impact on the $d$ quark PDF, especially in the small-$x$ range from 
$10^{-5}$ to $10^{-3}$. With a 300\invfb LHCb 13\tev $W^\pm$ data sample, the $d$ quark PDF uncertainty can be reduced by a factor of 30\%
around $x=10^{-3}$.
The LHCb 13\tev $W^\pm$ boson data have a smaller impact on the $u$ quark
PDF as compared to the $d$ quark PDF, but with a 300\invfb data sample, 
the LHCb 13\tev $W^\pm$ boson data do have an impact on the $u$ quark PDF in the small-$x$ region. 

\begin{figure}[!htbp]	
\centering
\includegraphics[width=0.45\textwidth]{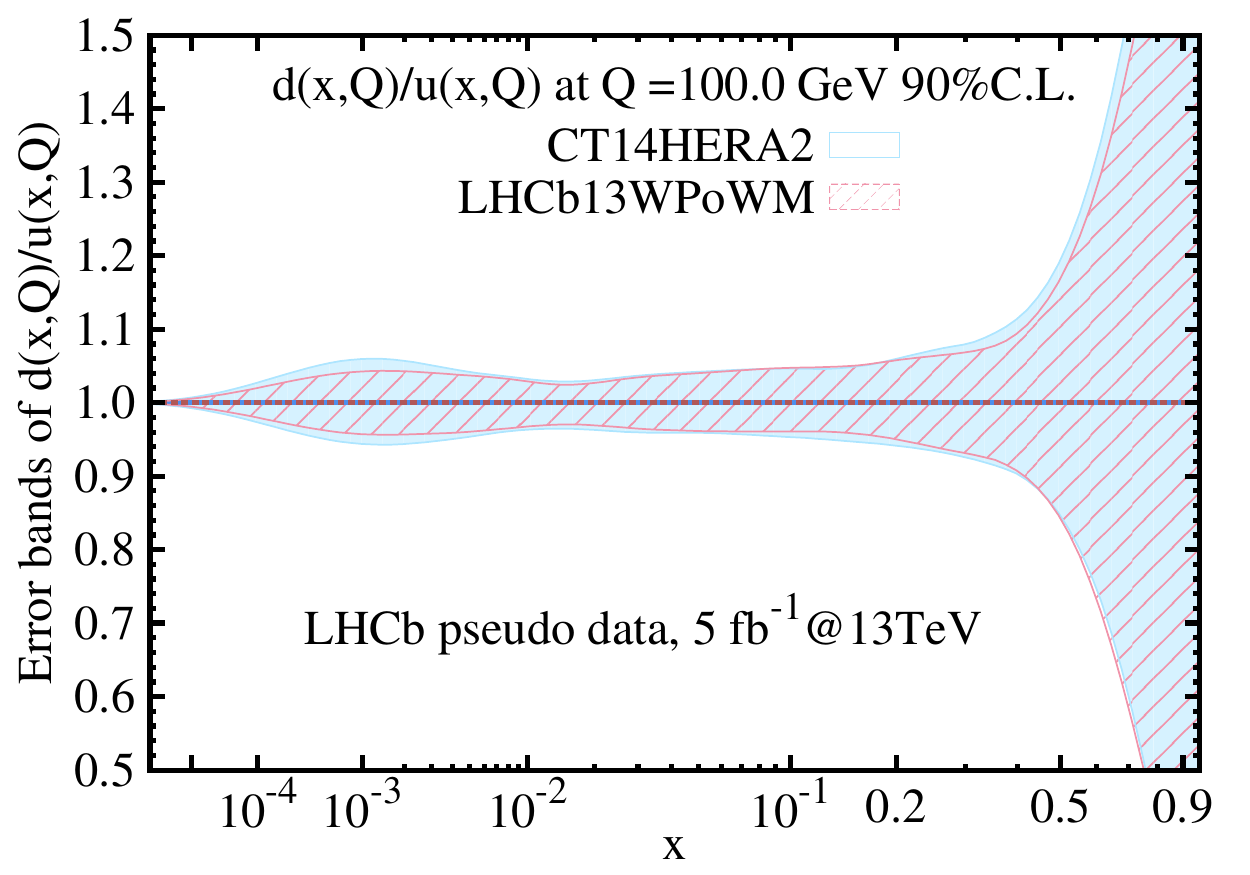}
\includegraphics[width=0.45\textwidth]{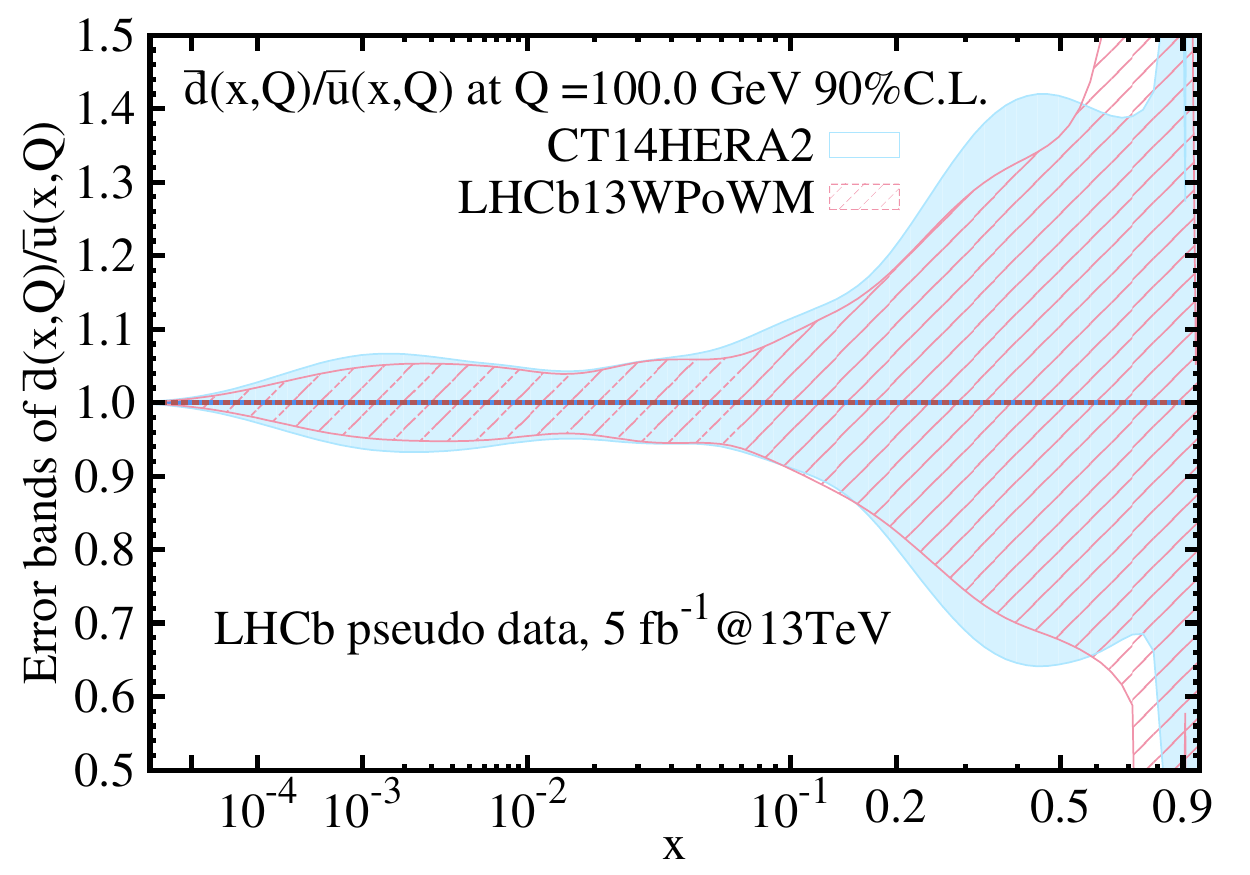}
\includegraphics[width=0.45\textwidth]{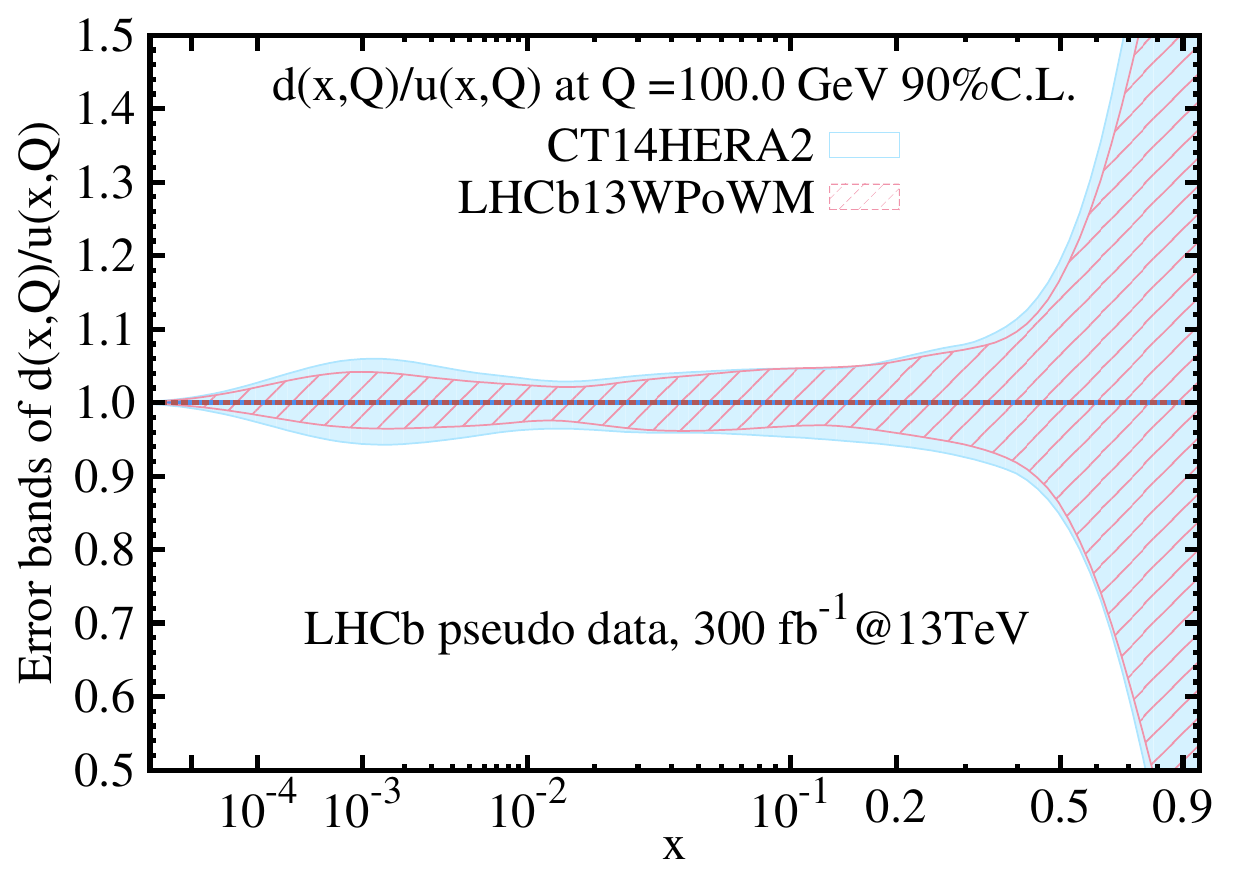}
\includegraphics[width=0.45\textwidth]{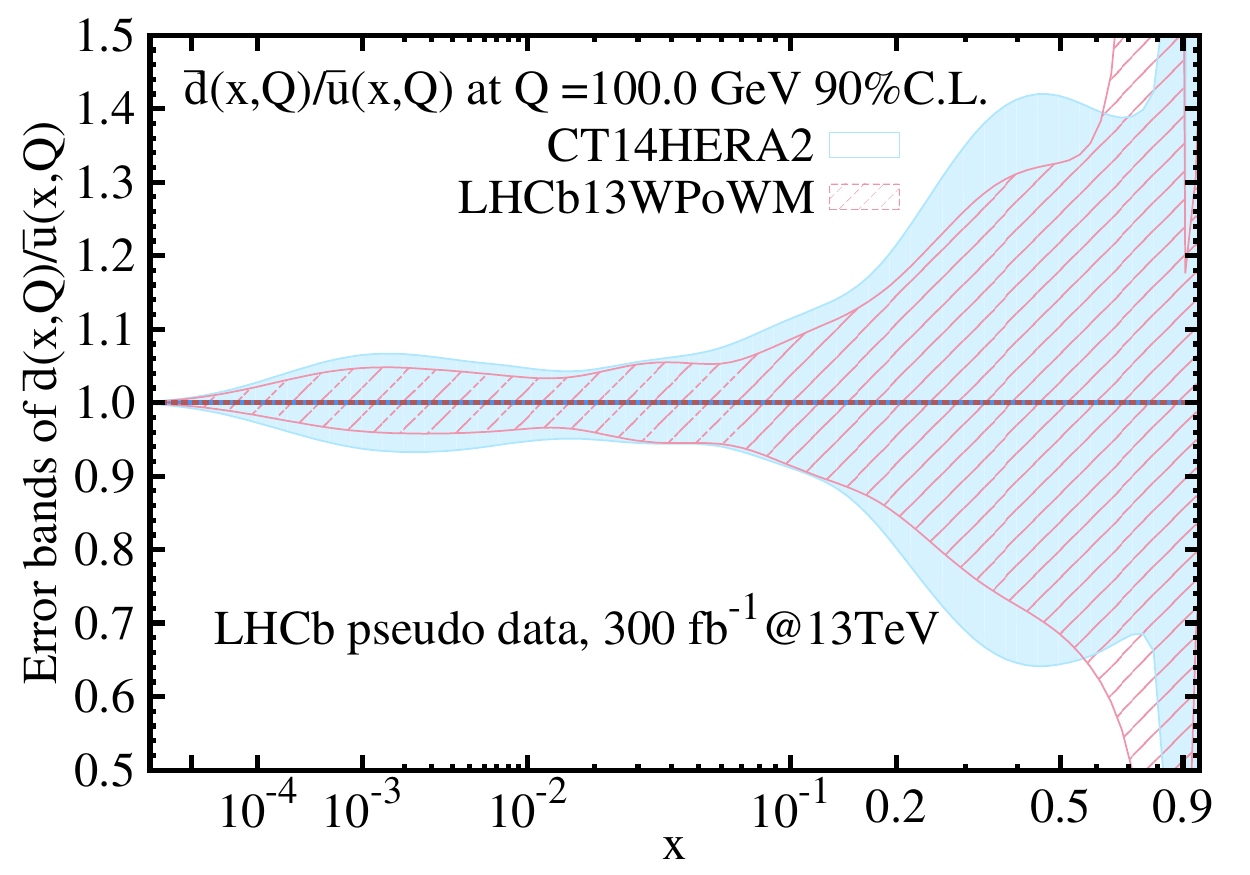}
\caption{
PDF uncertainties associated with \doveru (left) and
\doverubar (right), as a function of $x$, in the CT14HERA2 PDFs and the {\sc ePump} updated PDFs. 
The charged lepton pseudorapidity distributions of $W^+/W^-$ events are used as the input for the {\sc ePump} update; 
The \doveru result (top-left) and the \doverubar result (top-right) using a 5\invfb $W^+/W^-$ data sample;
the \doveru result (bottom-left) and the \doverubar result (bottom-right) using a 300\invfb $W^+/W^-$ data sample.
 }
\label{fig:udratio_wpm}
\end{figure}

The impacts of the LHCb 13\tev $W^+/W^-$ data, the ratio of $W^+$ and $W^-$ event rates, on the PDF ratios 
\doveru and \doverubar
are shown in Fig.~\ref{fig:udratio_wpm}. As we see, even with a 5\invfb data sample, the LHCb 13\tev $W^+/W^-$ data can already reduce the \doveru PDF uncertainty. 
Most of the improvements are concentrated in the small-$x$ region, from $10^{-5}$ to $10^{-3}$. 
The 300\invfb LHCb 13\tev data sample could further reduce the uncertainties of both the PDF ratios \doveru and \doverubar (by about $\sim 20\%$)
in the small-$x$ region, as well as giving some noticeable improvements in the large-$x$ region.
In the current PDF global fitting, the DIS data provide the largest constraint on the PDF ratio \doveru, cf. Ref.~\cite{Hou:2019gfw}. In the future, the LHCb data could provide additional information on the PDF ratios \doveru and \doverubar.

%%%%%%%%%%%%%%%%%%%%%%%%%%%%%%%%%%%%%%%%%%%%%%%%
\subsection{Update from LHCb 13 TeV $Z$ pseudo-data}\label{subsec:z}

The single $Z$ boson leptonic decay has two charged leptons in the final state.
These two charged leptons have large transverse momentum, and are isolated in the detector.
Based on these features, the $Z\rightarrow \ell^+\ell^-$ events are easy to
reconstruct and identify in a hadron collider, with small
background contamination.
Therefore, the $Z\rightarrow \ell^+\ell^-$ channel is one of best channels to 
perform precision EW measurements.

We used the $Z$ boson rapidity distribution as an observable for the {\sc ePump} update,
explore other observables that could be used in future PDF fitting, 
and propose a novel way to present $Z$ boson production measurement that provides more
valuable information for PDF fitting.
In this study, a binning scheme similar to the previous LHCb publication~\cite{Aaij:2016mgv} is used.

\begin{figure}[!htbp]	
\centering
 \includegraphics[width=0.45\textwidth]{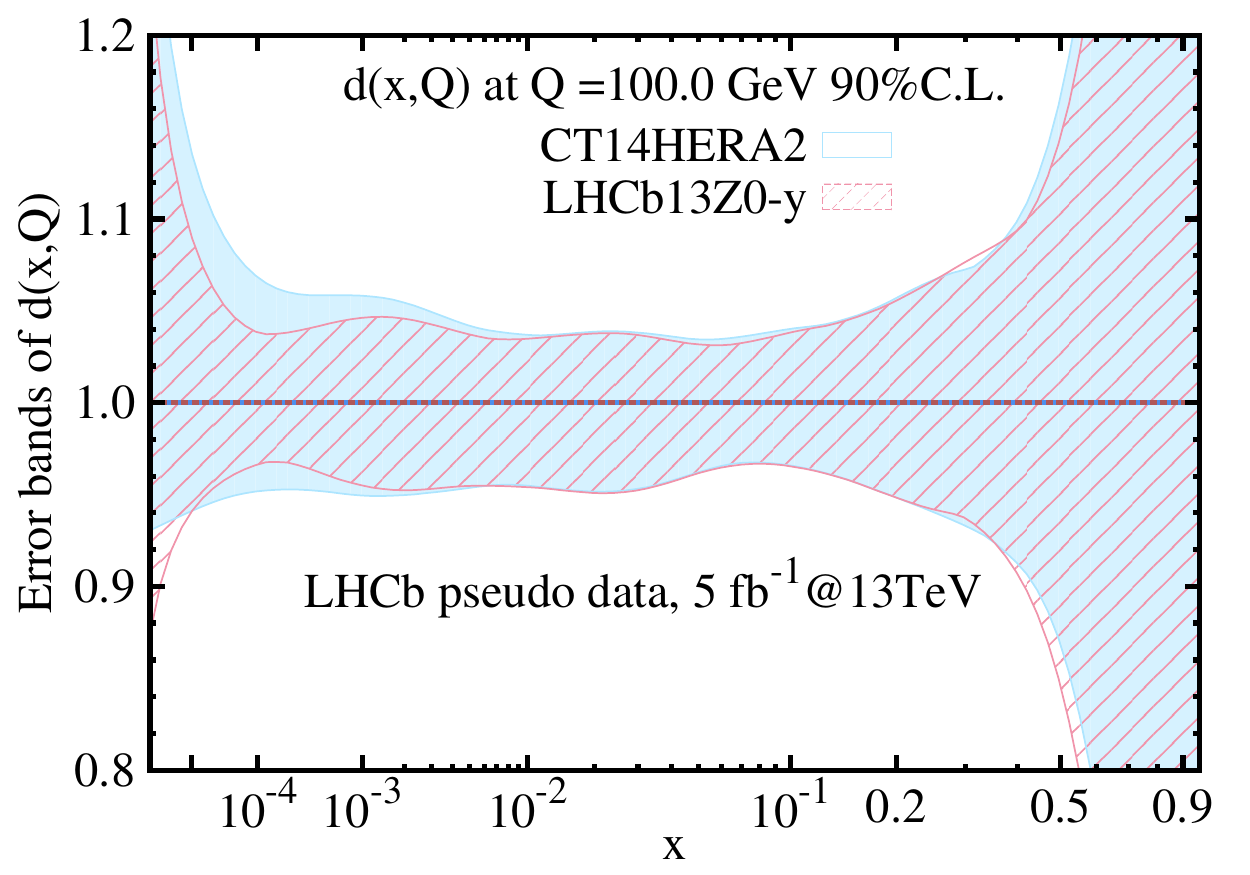}
 \includegraphics[width=0.45\textwidth]{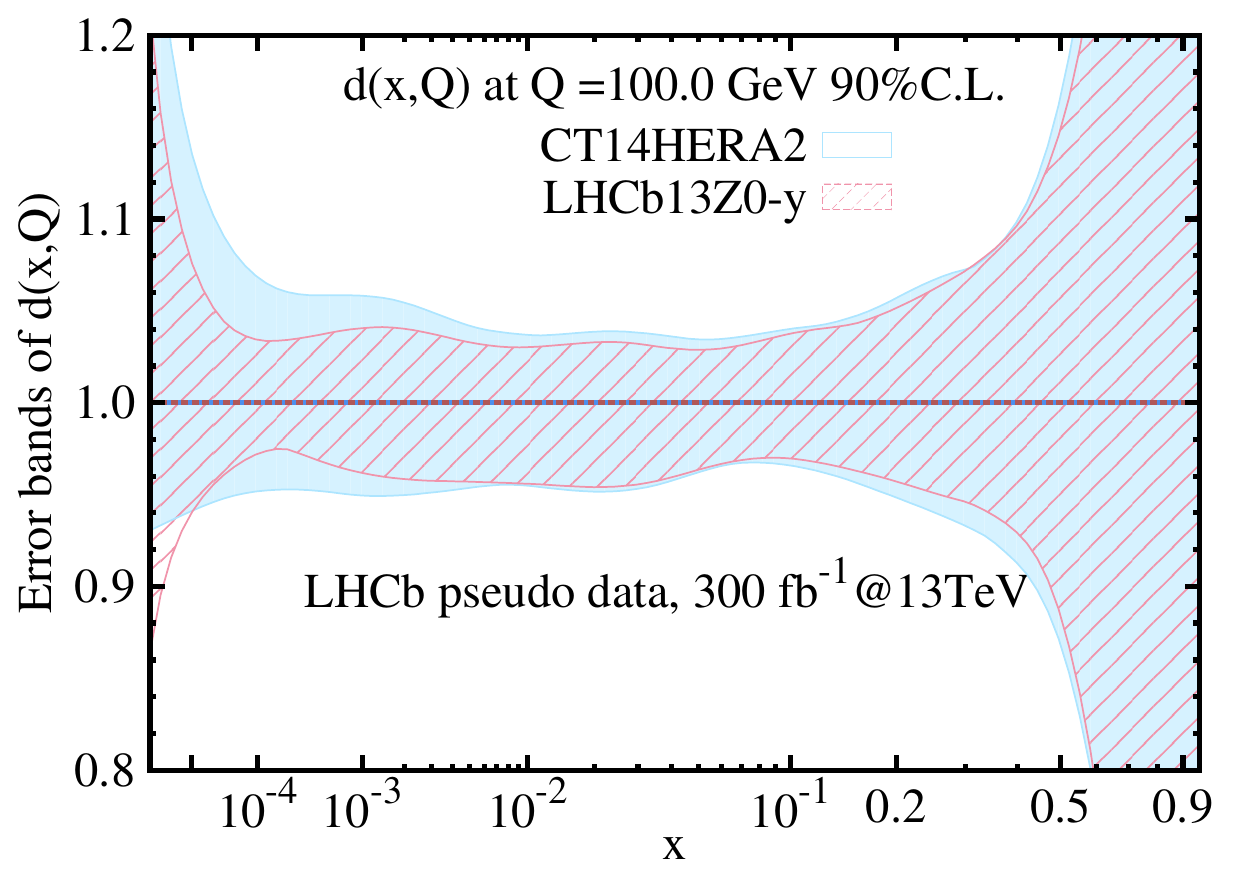}
 \caption{PDF uncertainties associated with the $d$ quark, as a function of $x$, 
in the CT14HERA2 PDFs and {\sc ePump} updated new PDFs. 
The denominator is the central value of each PDF set.
The rapidity distribution of $Z$ boson events is used as input of the {\sc ePump} update. 
(left) The $d$ quark result using 5\invfb data, and (right) the $d$ quark result using 300\invfb data.}
\label{fig:z_udquark}
\end{figure}

The updated PDF results are shown in Fig.~\ref{fig:z_udquark} for the $d$ quark. As shown in the figure, with 5\invfb of data sample, the LHCb 13\tev single $Z$ boson data is not as powerful as $W^{\pm}$ data, mainly due to its smaller event rate. 
With 300\invfb, however, the $Z$ boson data has an impact on $d$ quark PDFs in the small-$x$ region, from $10^{-5}$ to $10^{-2}$.

\begin{figure}[!htbp]	
\centering
\includegraphics[width=0.45\textwidth]{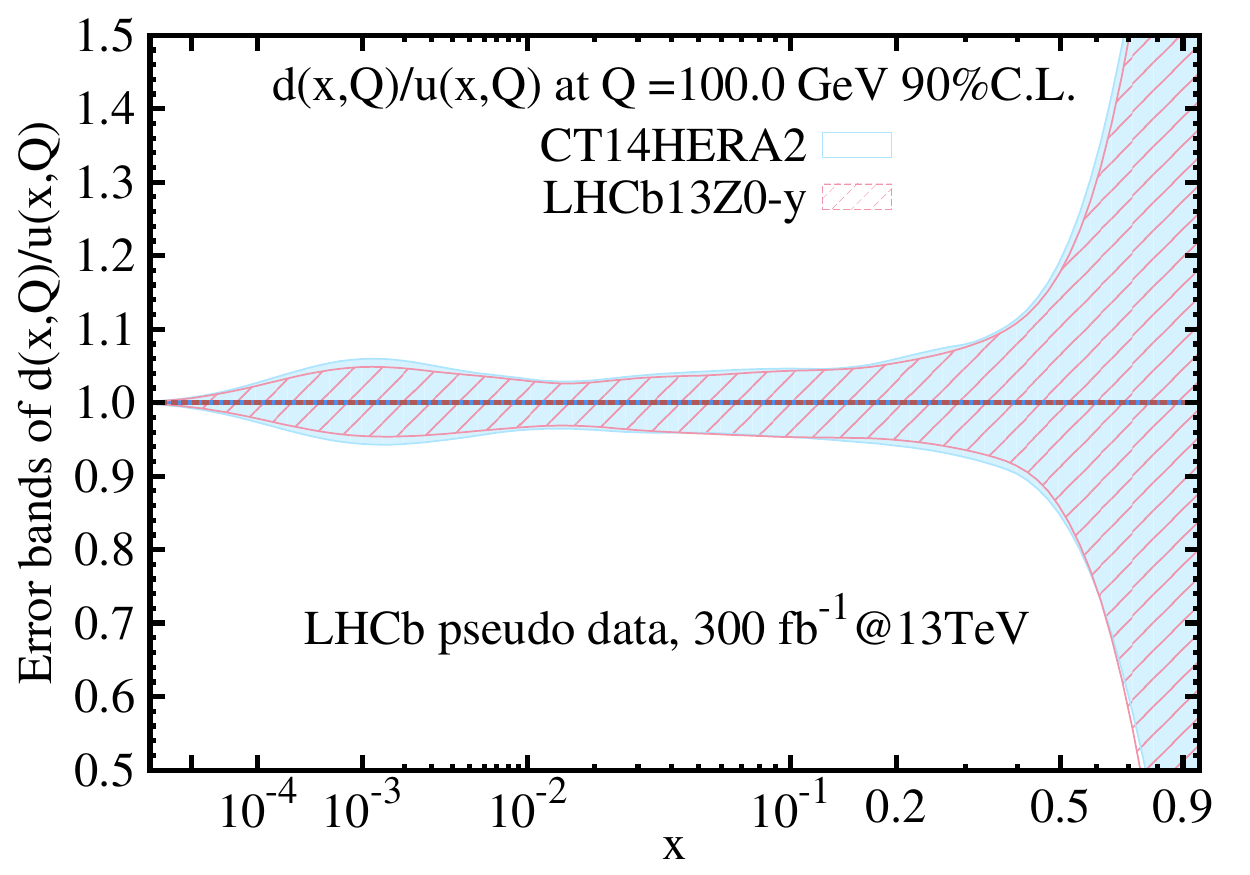}
\includegraphics[width=0.45\textwidth]{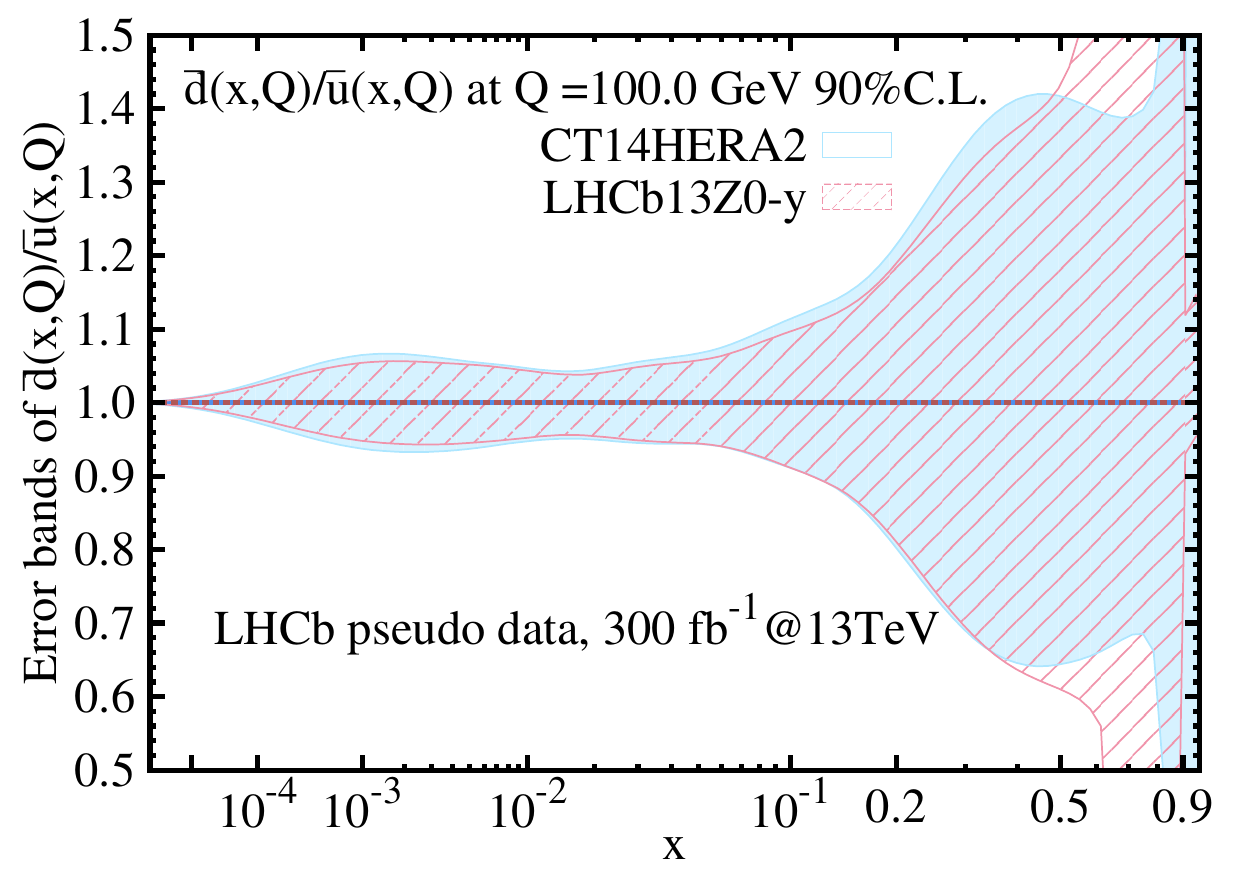}
\caption{PDF uncertainties associated with \doveru (left) and
\doverubar (right), as a function of $x$, 
in the CT14HERA2 PDFs and {\sc ePump} updated new PDFs. 
The denominator is the central value of each PDF set.
The rapidity distribution of $Z$ boson events is used as input of the {\sc ePump} update. 
(left) The \doveru quark result using 300\invfb data,
and (right) the \doverubar quark result using 300\invfb data.}
\label{fig:z_udratioquark}
\end{figure}

The impacts from the LHCb 13\tev 300\invfb single $Z$ boson data on \doveru and \doverubar are shown in Fig.~\ref{fig:z_udratioquark}, 
where the LHCb $Z$ boson data could reduce the \doveru PDF uncertainty in the small-$x$ region. 

\begin{figure}[!htbp]	
\centering
\includegraphics[width=0.45\textwidth]{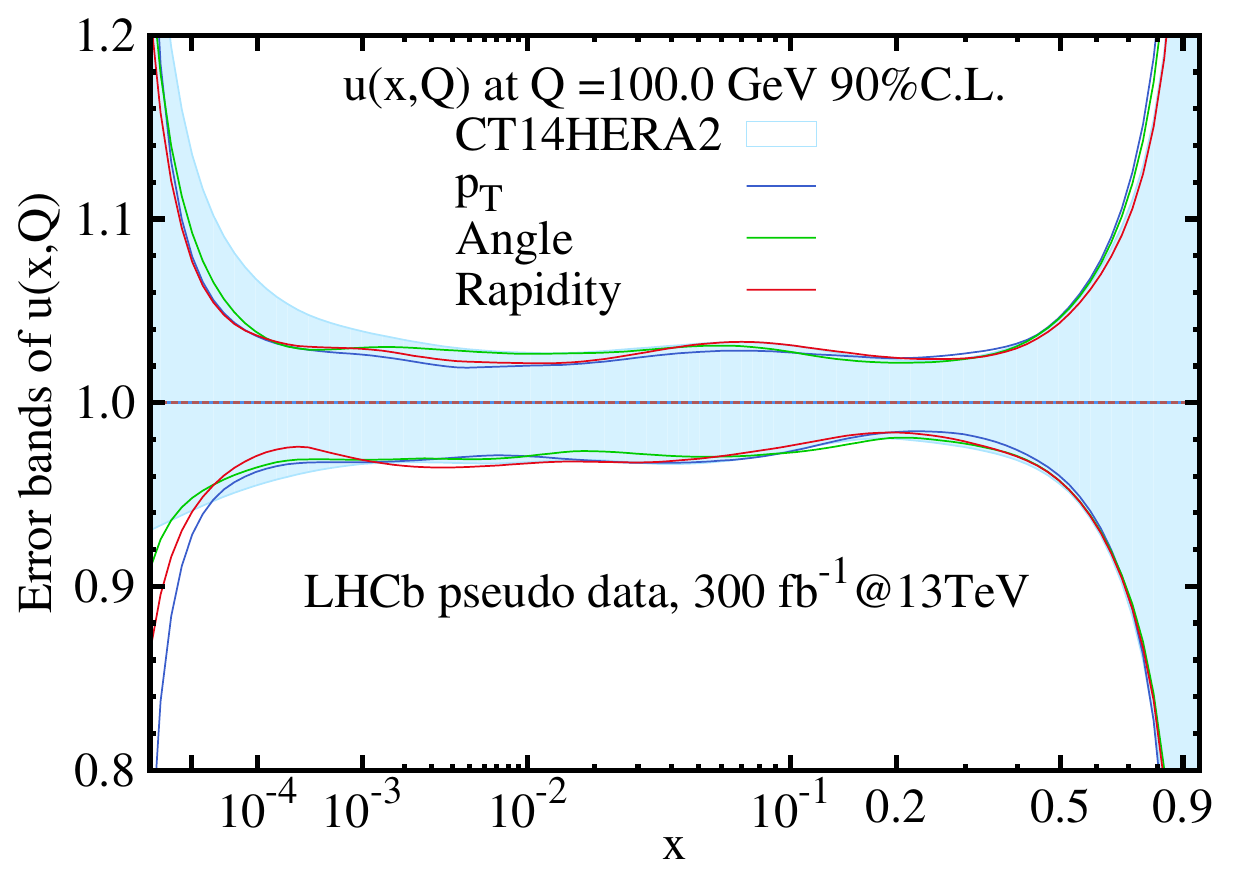}
\includegraphics[width=0.45\textwidth]{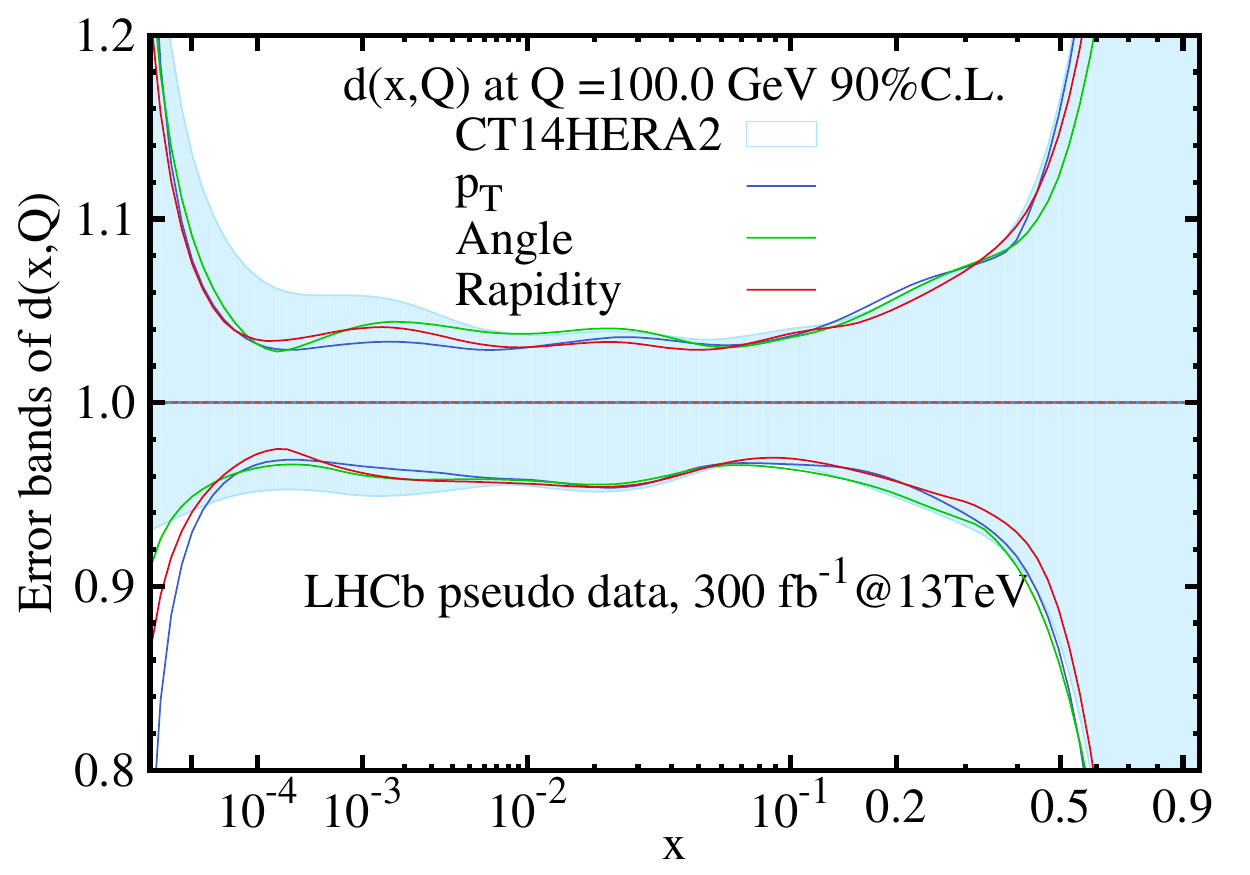}
\caption{
PDF uncertainties associated with the $u$ quark (left) and $d$ quark (right), as a function of $x$, in the CT14HERA2 PDFs and {\sc ePump} updated new PDFs. The denominator is the central value of each PDF set.
The single $Z$ boson events (300\invfb) are used in the {\sc ePump} update.}
\label{fig:Zvarscheck}
\end{figure}

We have also explored the sensitivity of 
the $Z$ boson $p_T$, lepton $\cos\theta^*$ (defined in the Collins-Soper frame~\cite{Collins:1977iv}), and $Z$ boson rapidity distributions measured at the LHCb to further constrain the PDFs.
We consider their impacts one at a time in the {\sc ePump} update. 
As shown in Fig.~\ref{fig:Zvarscheck}, 
we see that each observable has a slightly different impact on the $u$ ($d$) quark PDFs across the whole $x$ region, as expected. 
 
\begin{figure}[!htbp]
\centering
\includegraphics[width=0.45\textwidth]{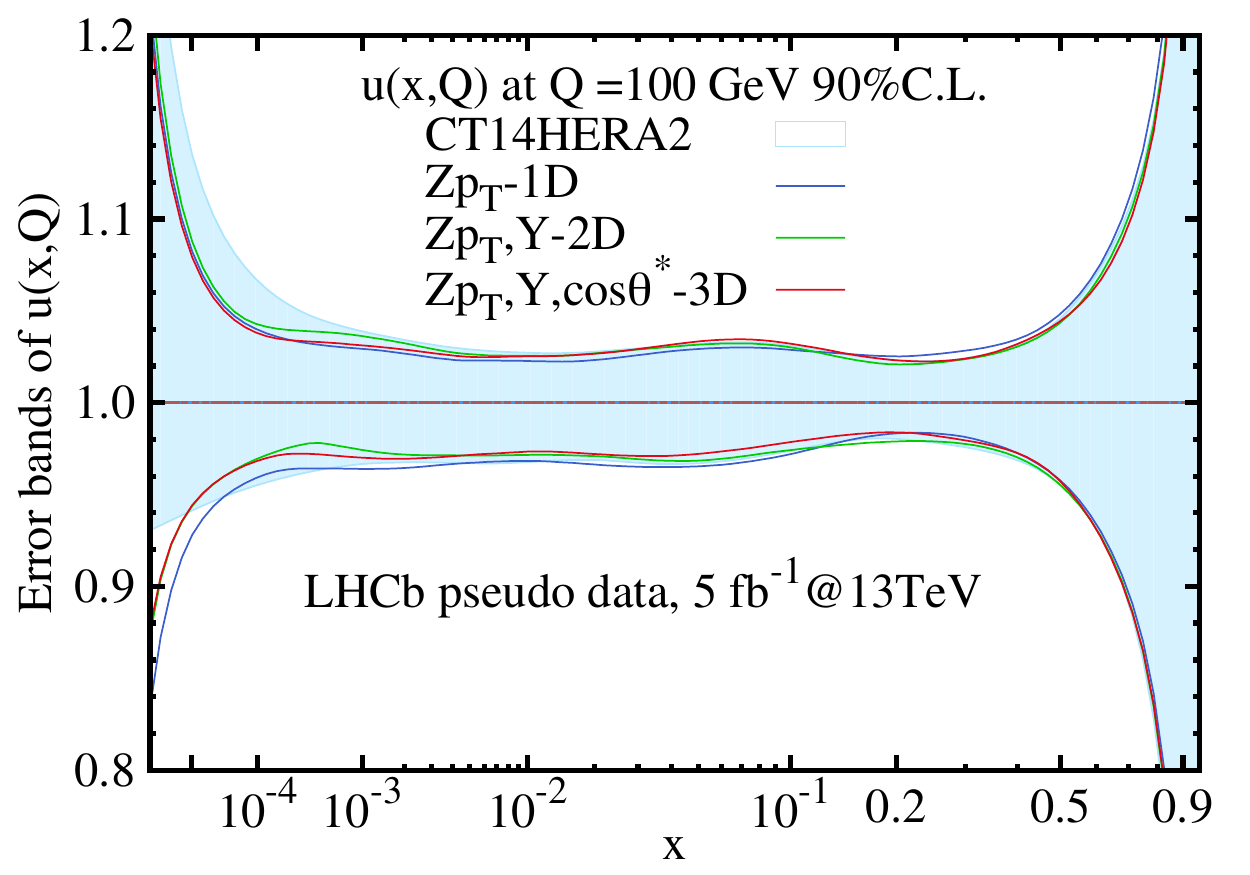}
\includegraphics[width=0.45\textwidth]{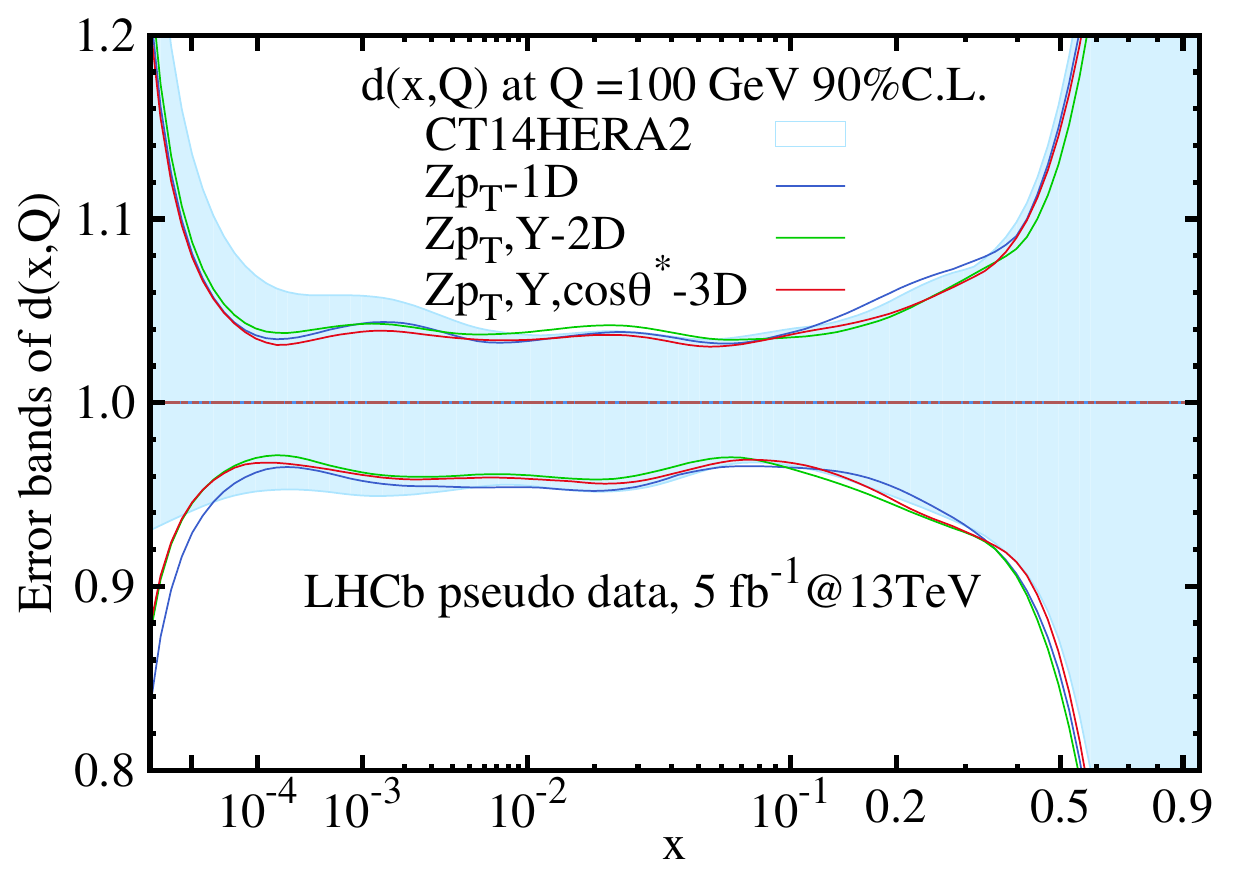}
\includegraphics[width=0.45\textwidth]{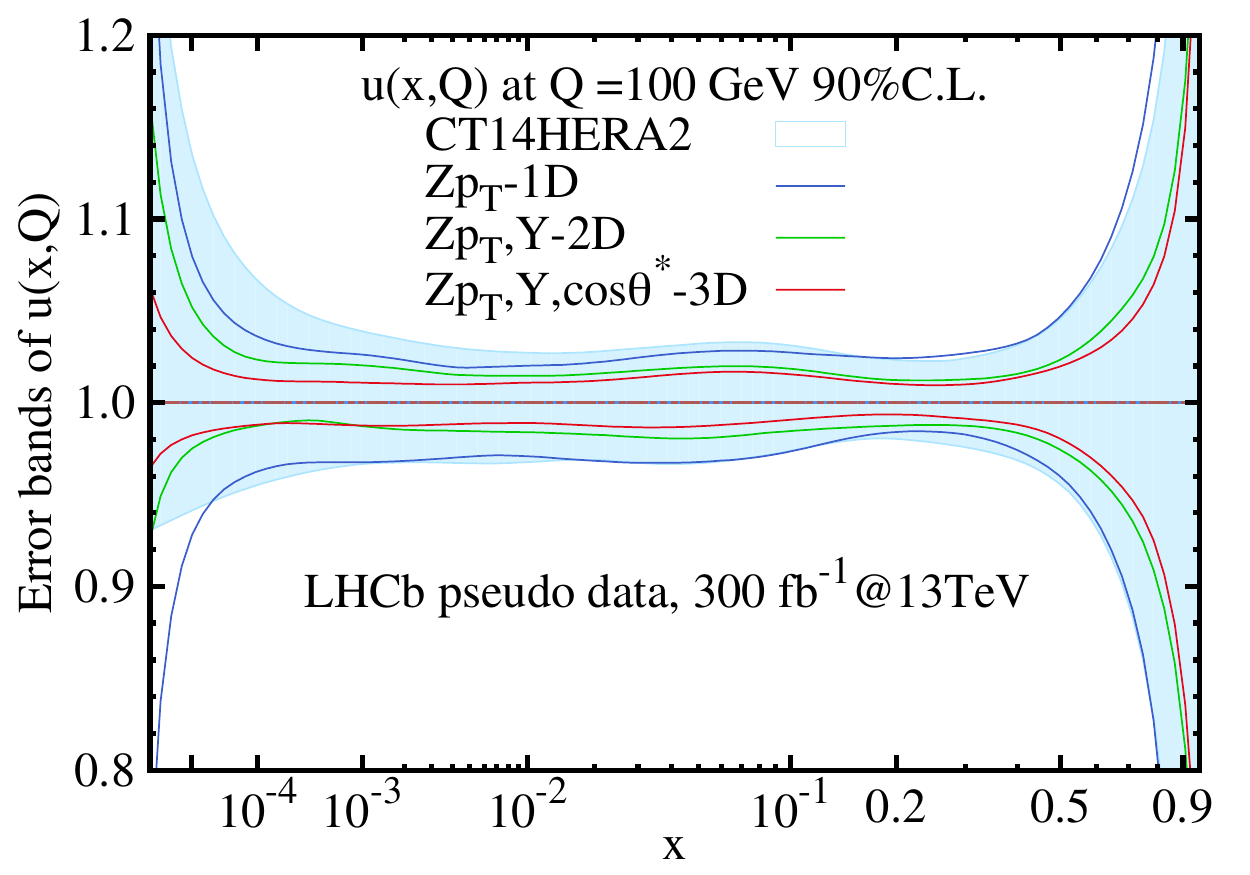}
\includegraphics[width=0.45\textwidth]{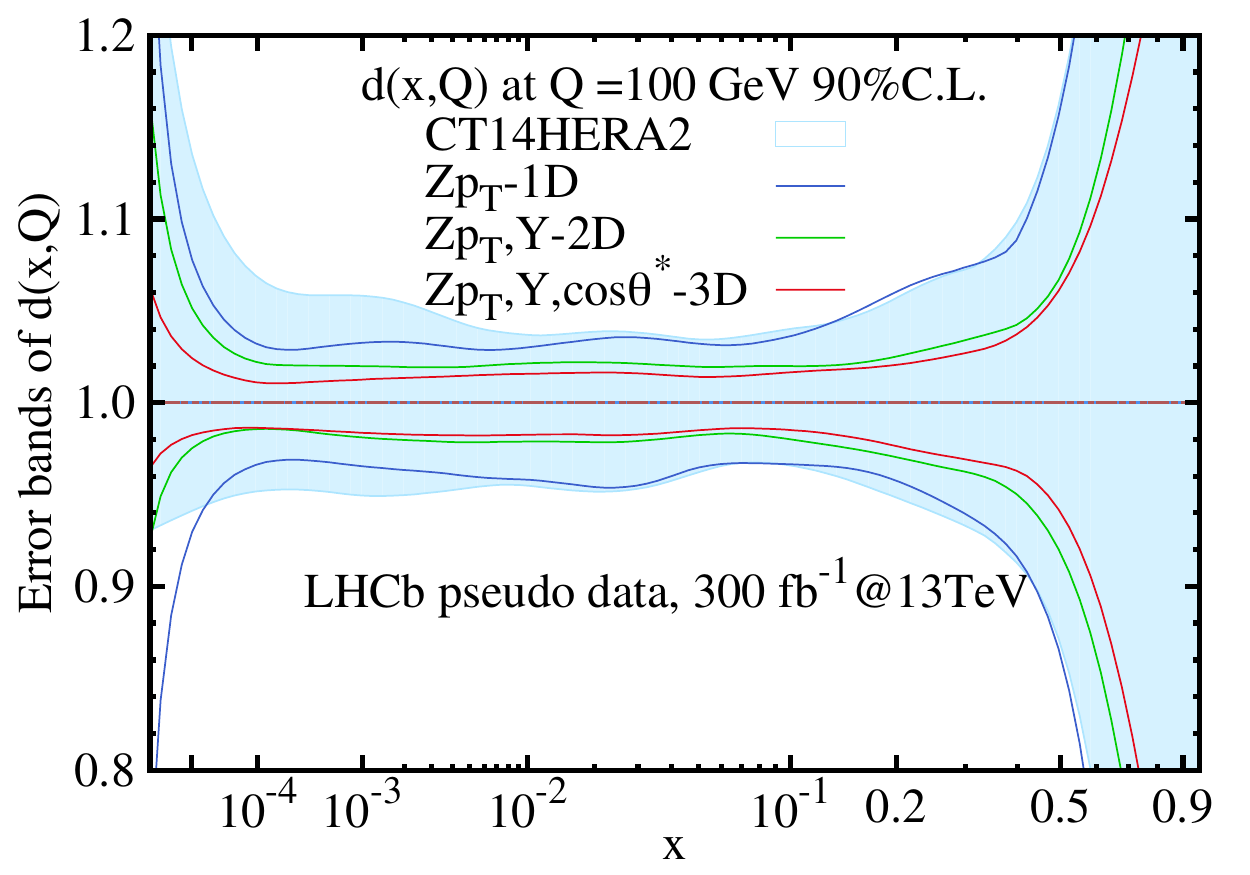}
\caption{PDF uncertainties associated with the $u$ quark (left) and
$d$ quark (right), as a function of $x$, 
in the CT14HERA2 PDFs and {\sc ePump} updated new PDFs. 
The denominator is the central value of each PDF set.
The single $Z$ boson multi-dimensional (1D, 2D, and 3D) differential cross section is used in the {\sc ePump} update.
(top-left) The $u$ quark result using 5\invfb data;
(top-right) the $d$ quark result using 5\invfb data;
(bottom-left) the $u$ quark result using 300\invfb data;
and (bottom-right) the $d$ quark result using 300\invfb data.}
\label{fig:1Dor3D}
\end{figure}

Below, we propose a better way to extract useful information from the LHCb 13\tev $Z$ data, by performing a multi-dimensional analysis. 
With more $pp$ collision data to be collected by the LHCb detector in the future, it is feasible 
to perform $Z$ boson production cross section measurement with 
a multi-dimension binning, like double- or triple-differential cross section measurements. 
Comparisons of updated PDF uncertainties with different numbers of (input) experimental observables are shown in Fig.~\ref{fig:1Dor3D}. 
As shown in the figure, we compare the impacts of the LHCb 13\tev $Z$ boson pseudo-data on PDFs by performing 
a single differential ($Z$ boson \pt, labeled as `1D' in the figure), 
a double-differential ($Z$ boson \pt and $Z$ boson rapidity, labeled as `2D' in the figure),
and a triple-differential ($Z$ boson \pt, $Z$ boson rapidity, and lepton $\cos\theta^*$, labeled as `3D' in the figure) 
cross section measurements.
We find that with limited statistics of $Z$ boson events (5\invfb data), 
the mulit-dimensional measurements cannot significantly improve the PDF determination, as compared to one-dimensional measurements on 
$Z$ boson \pt, lepton $\cos\theta^*$, and $Z$ boson rapidity, respectively. 
With a 300\invfb data sample, however, the mulit-dimensional measurement has 
better constraints on the PDFs, across the whole $x$ range.
The triple-differential cross section gives the best constraints on $u$ and $d$ quark PDFs, across the whole $x$ range. 
The improvements gained in going from `2D' to `3D' measurement is not as
strong as that from `1D' to `2D' measurement. 
From the experimental point of view, a triple-differential cross section measurement could have limited statistics 
in extreme phase space, such as at the boundaries of observables. Furthermore, it is complicated to calculate 
correlated systematic uncertainties in a `3D' measurement, compared to a `2D' measurement. 
Therefore, with large future data samples, a double-dimensional $Z$ boson cross section measurement (a double-differential $Z$ boson \pt and $Z$ boson rapidity)
is feasible and recommended, and could provides more valuable information in PDF fitting than a single- or a triple-dimensional measurement.

In the double- and triple-differential cross section measurements, the 
binning schemes for $Z$ boson $\pt$, rapidity ($y_{\ell\ell}$), and lepton $\cos\theta^*$ are defined as:

\begin{itemize}
\item $0<\pt<250$: \{0, 1, 2, 3, 4, 5, 6, 7, 8, 9, 10, 11, 12, 13, 14, 15, 17.5, 20, 22, 25, 28, 33, 40, 50, 100, 250\}\gevc.
\item $2<y_{\ell\ell}<4.5$: \{2.00, 2.14, 2.28, 2.42, 2.56, 2.69, 2.83, 2.97, 3.11, 3.25, 3.39, 3.53, 3.67, 3.81, 3.94, 4.08, 4.22, 4.36, 4.50\}.
\item $-1<\cos\theta ^*<1$: \{-1, 0, 1\}.
\end{itemize}

%%%%%%%%%%%%%%%%%%%%%%%%%%%%%%%%%%%%%%%%%%%%%%%%
\subsection{Update from LHCb 13 TeV $W^+ \;+\; W^- \;+\; Z$ pseudo-data}
\label{subsec:Combination}

In general, global fitting of PDFs should use all available experimental data as inputs. Therefore, we checked the impact from the LHCb 13\tev data on the PDF fitting, including both single $W^{\pm}$ boson and $Z$ boson data samples. 
In reality, as $W^{\pm}$ and $Z$ boson results from one experiment are measured with the same data sample, many systematic uncertainties are correlated.
In any PDF fitting, correlation matrices between different observables must be provided to avoid potential data bias. 
In this study, without detector-level simulated events, we cannot calculate the 
correlation matrices between the single $W^{\pm}$ and $Z$ boson measurements.
So we assume in this study that there is no correlation
between the LHCb 13\tev $W^{\pm}$ and $Z$ pseudo-data.
In the study, the $W^{\pm}$ and $Z$ boson single differential cross section results are used as the inputs of the {\sc ePump} update, which are 
the charged lepton $\eta$ distribution of $W^{\pm}$ boson events and the rapidity distribution of $Z$ bosons.

\begin{figure}[!htbp]	
\centering
 \includegraphics[width=0.45\textwidth]{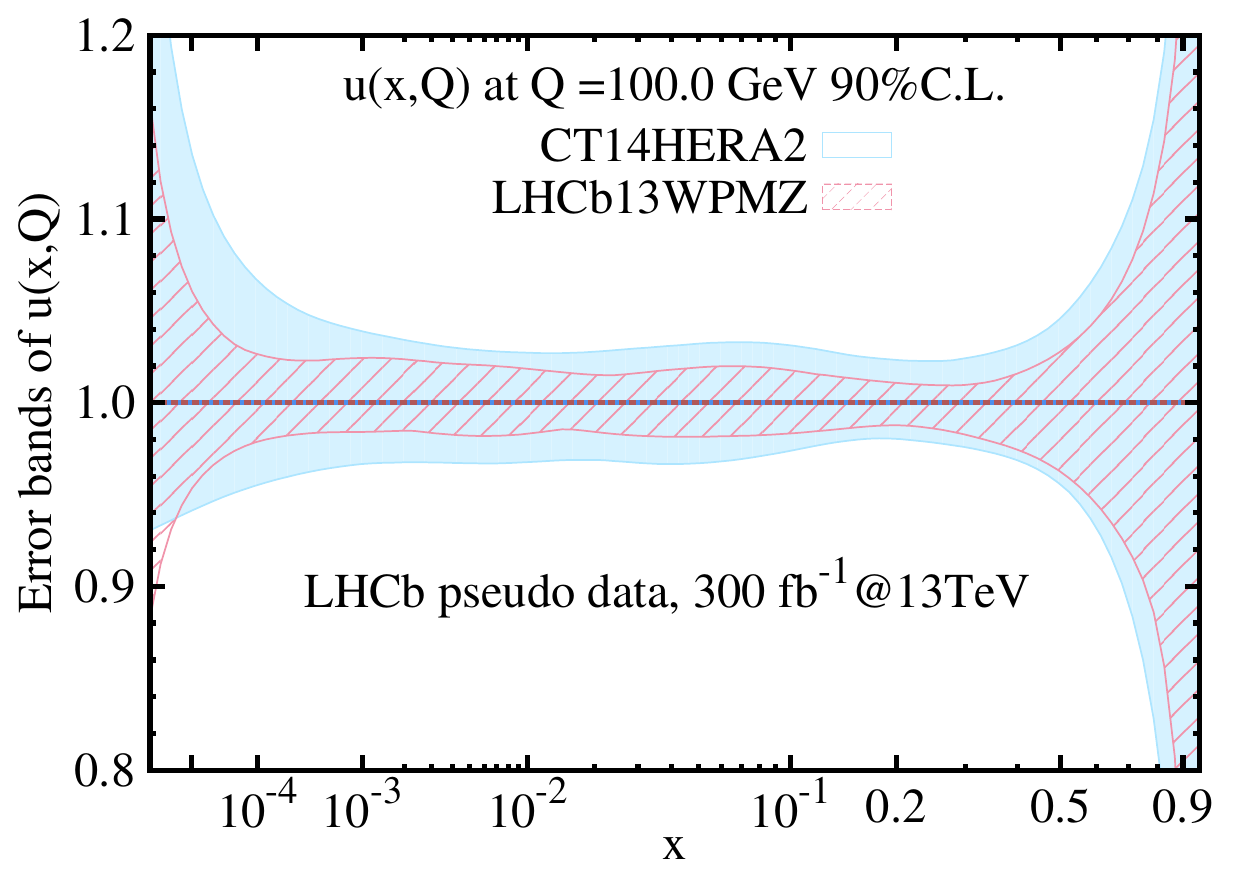}
 \includegraphics[width=0.45\textwidth]{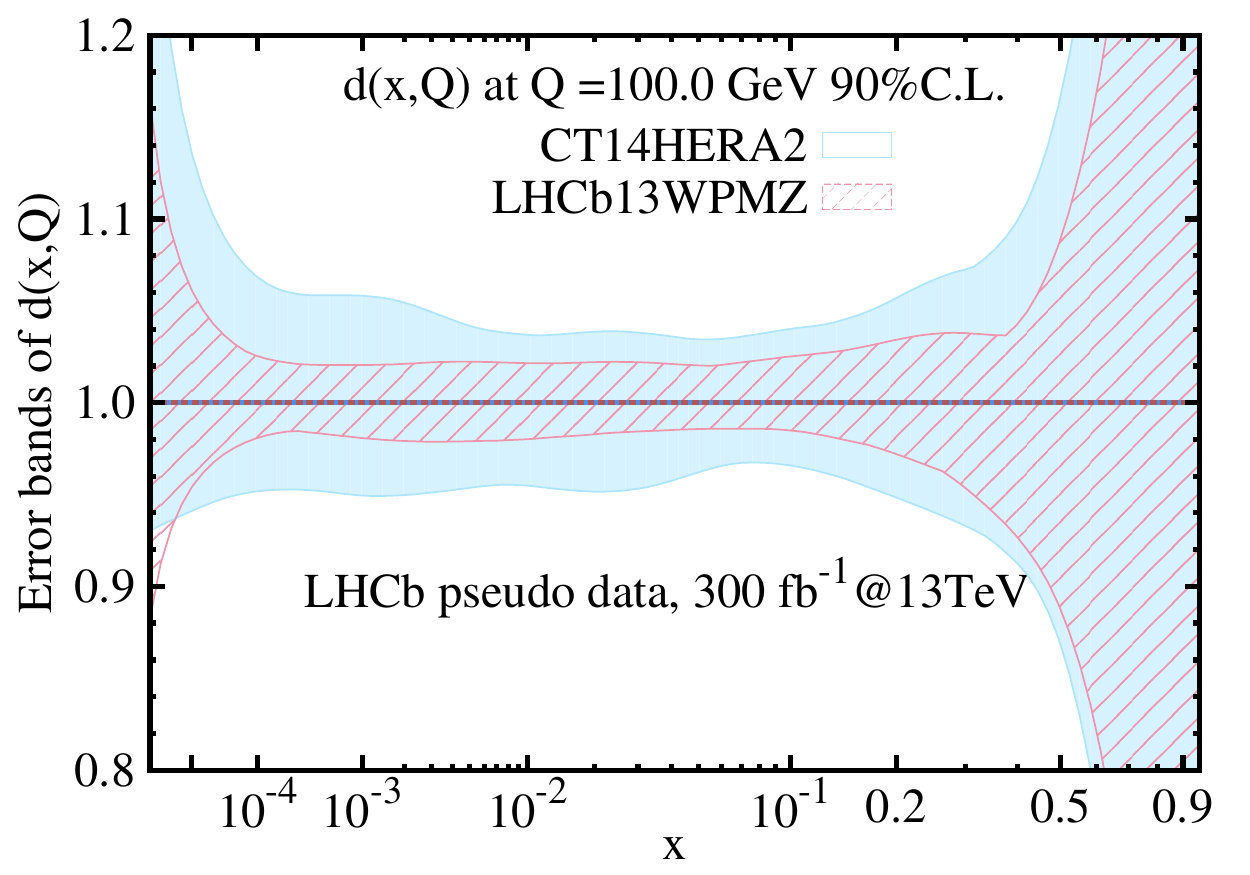}
\includegraphics[width=0.45\textwidth]{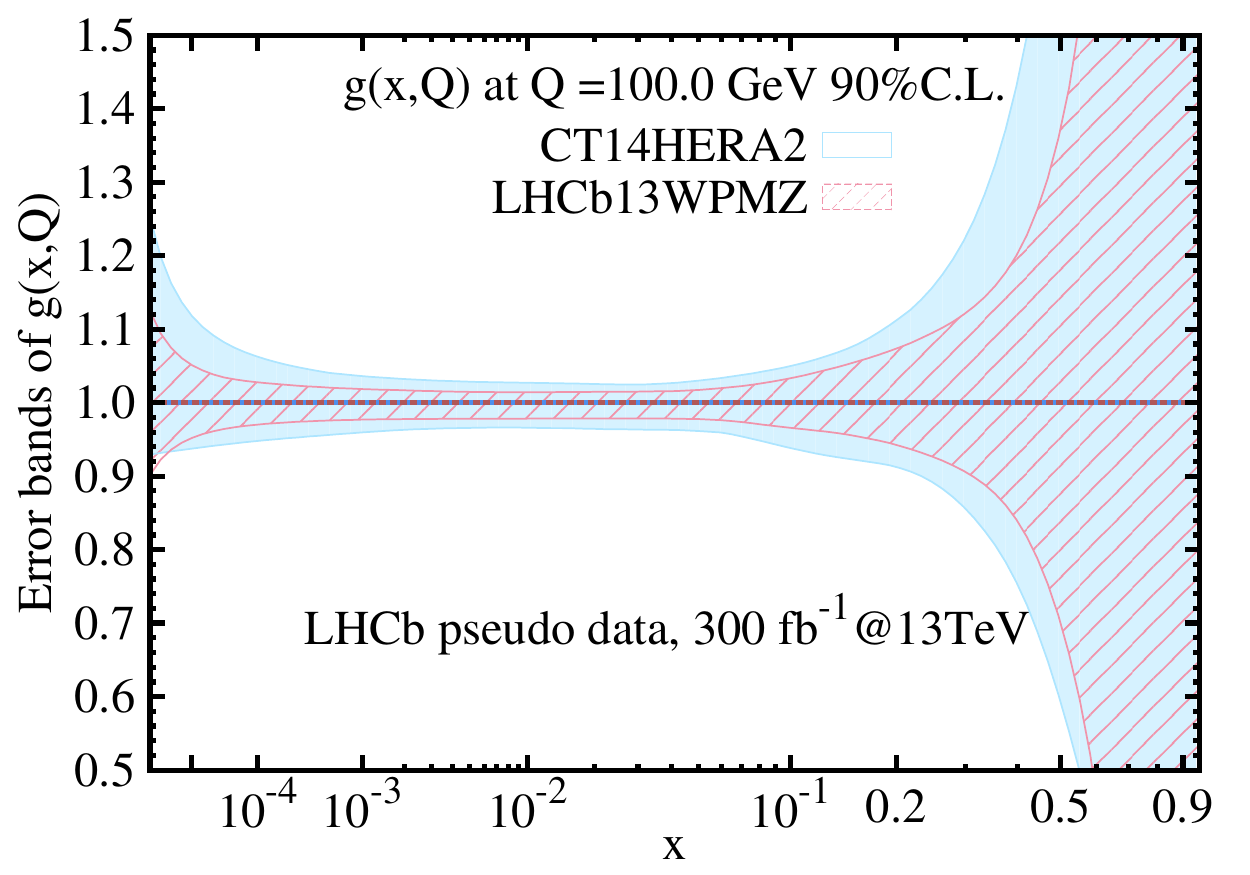}
\includegraphics[width=0.45\textwidth]{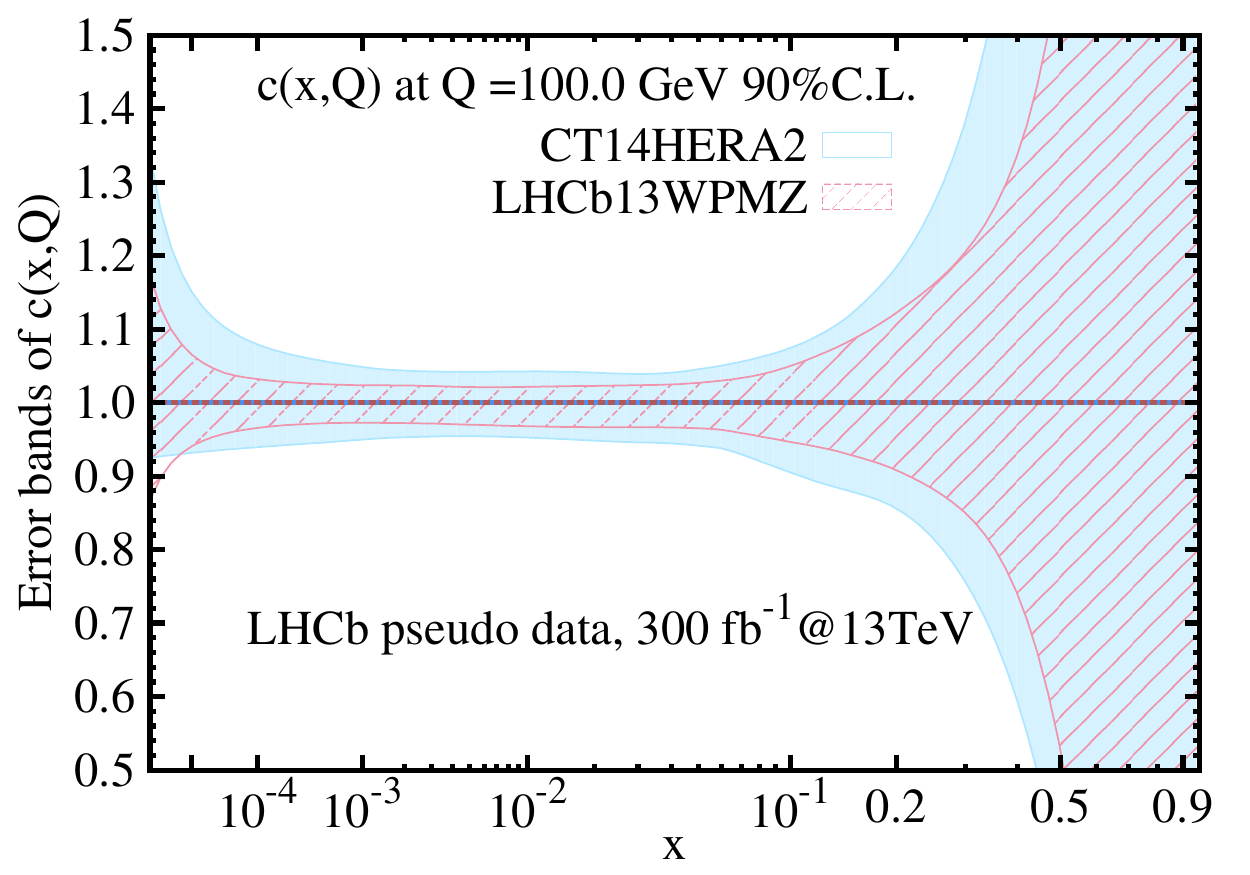}
\caption{PDF uncertainties associated with the $u$ quark (left) and
$d$ quark (right), as a function of $x$, 
in the CT14HERA2 PDFs and {\sc ePump} updated new PDFs. 
The denominator is the central value of each PDF set.
The 300\invfb single $W^{\pm}$ and $Z$ boson differential cross section are taken as {\sc ePump} input. 
(top-left) The $u$ quark result; (top-right) the $d$ quark result; (bottom-left) the gluon result; and (bottom-right) the $c$ quark result.}
\label{fig:wzinputs_ud}
\end{figure}

\begin{figure}[!htbp]	
\centering
\includegraphics[width=0.45\textwidth]{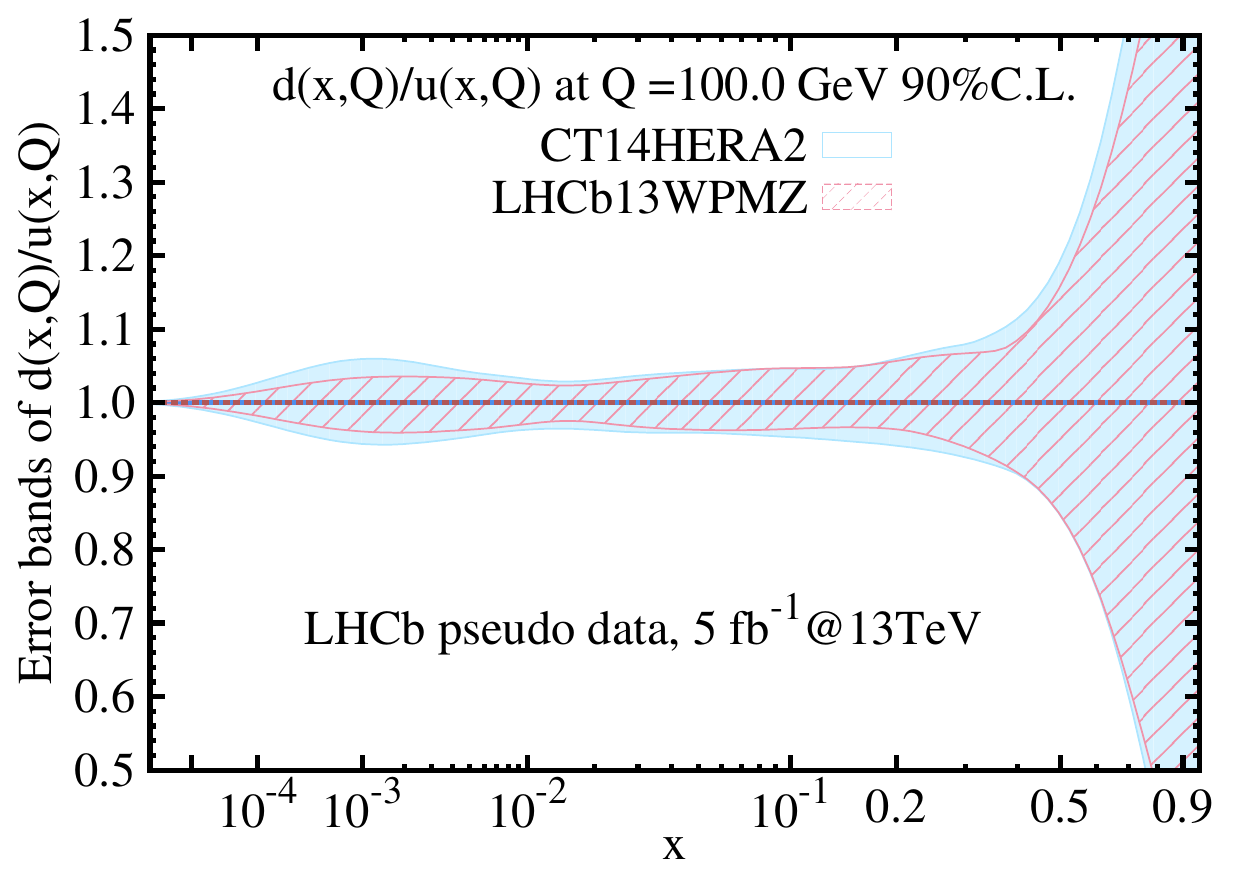} 
 \includegraphics[width=0.45\textwidth]{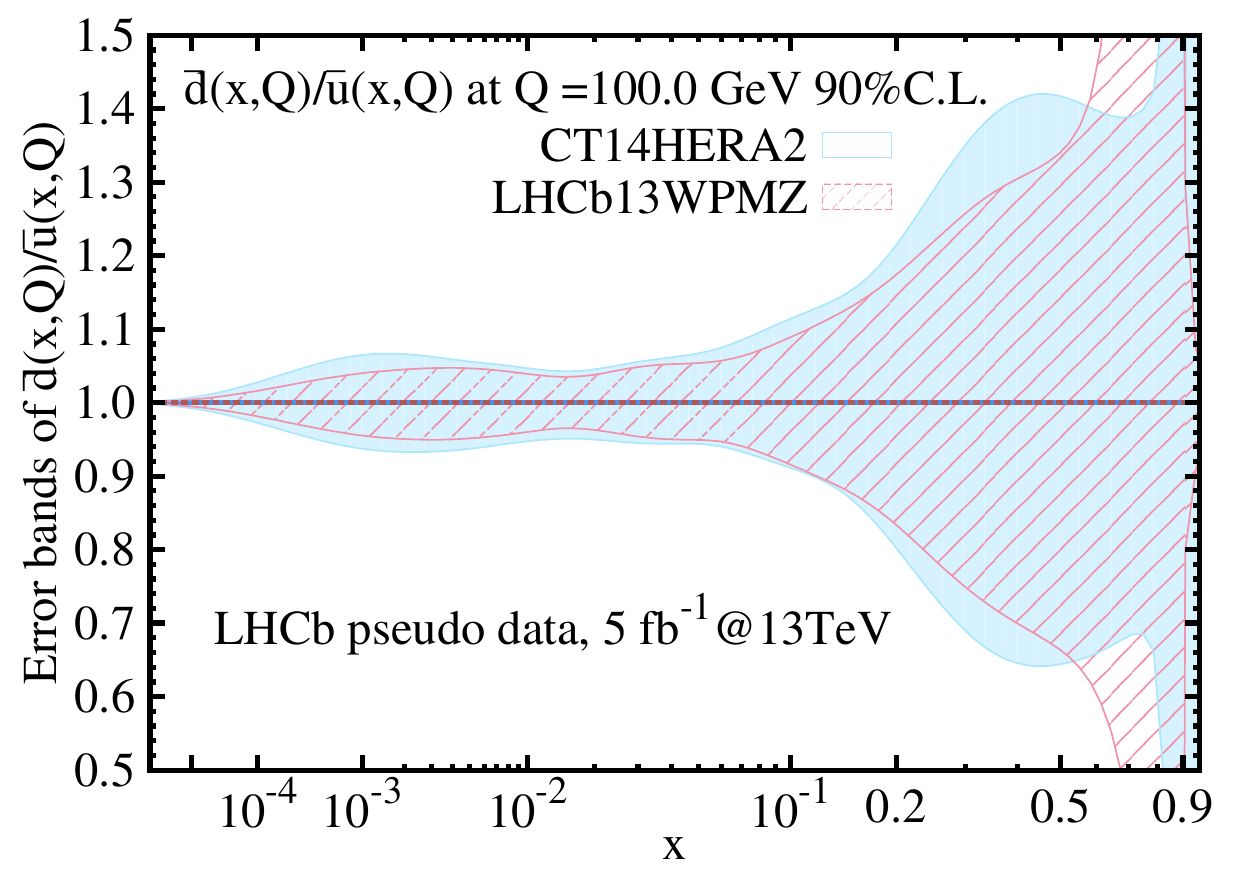}
 \includegraphics[width=0.45\textwidth]{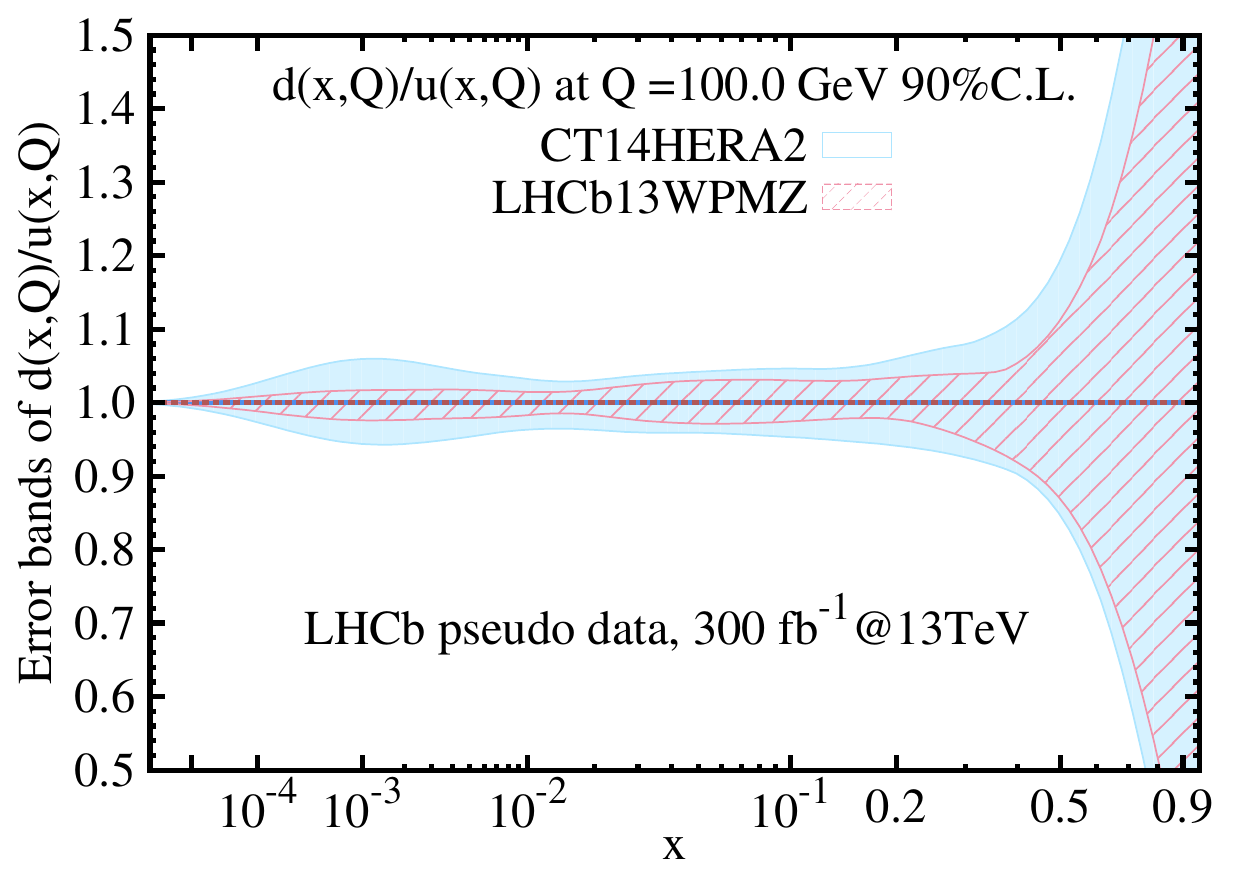}
 \includegraphics[width=0.45\textwidth]{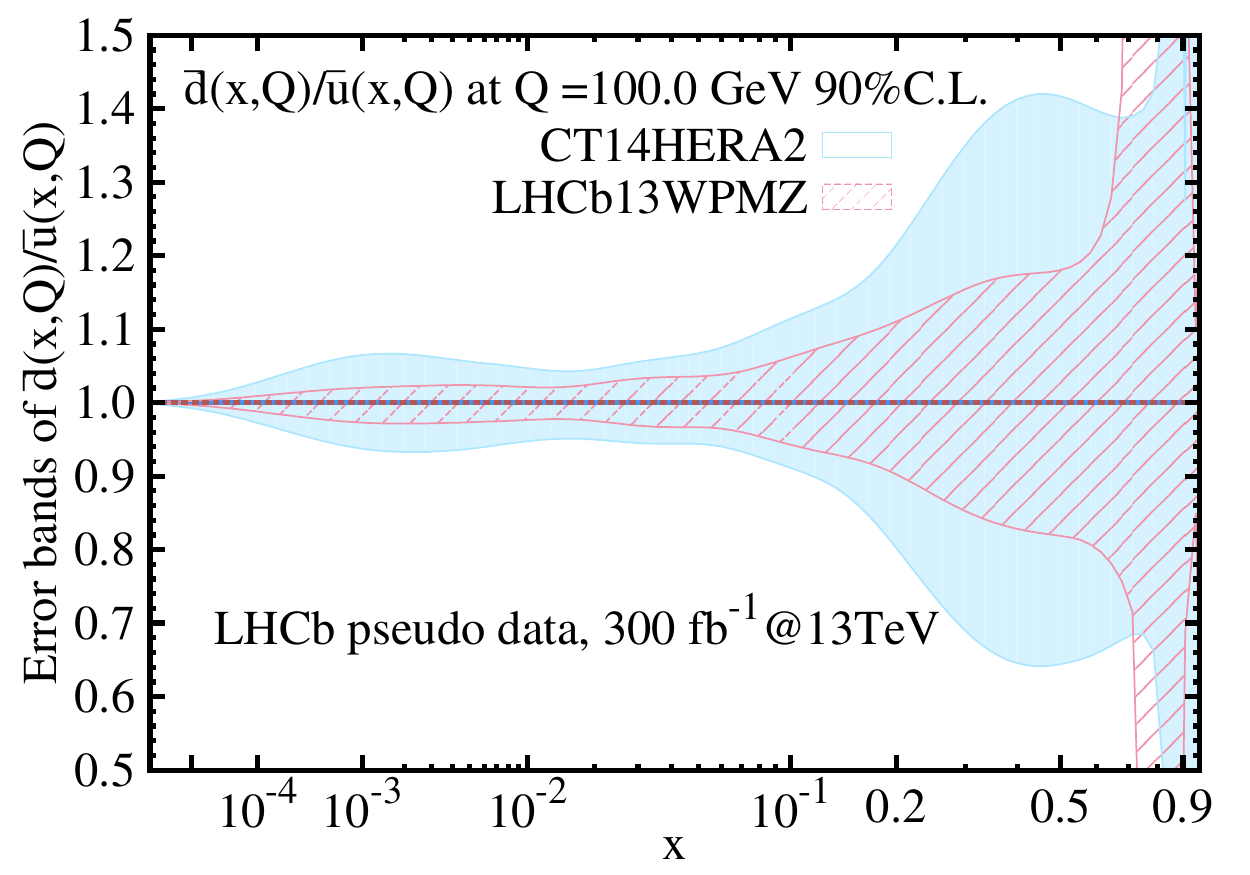}
 \caption{PDF uncertainties associated with \doveru (left) and \doverubar (right), as a function of $x$, in the CT14HERA2 PDF sets
and {\sc ePump} updated new PDF sets. The denominator is the central value of each PDF set.
The single $W^{\pm}$ and $Z$ boson differential cross section are taken as {\sc ePump} input.
(top-left) The \doveru result using 5\invfb data;
(top-right) the \doverubar result using 5\invfb data;
(bottom-left) the \doveru result using 300\invfb data;
and (bottom-right) the \doverubar result using 300\invfb data.}
\label{fig:wz_inputs_udratio}
\end{figure}

The updated PDF results are shown in Fig.~\ref{fig:wzinputs_ud} for $u$, $d$, $c$ quark and gluon PDFs,
and the \doveru and $\bar{d}/\bar{u}$ ratio results are shown in 
Fig.~\ref{fig:wz_inputs_udratio}. 
Based on these figures, the following features are found:
\begin{itemize}
\item The largest improvement is in the $d$ quark PDFs. The uncertainty of the $d$ quark PDFs 
 can be improved significantly by the LHCb 13\tev $W^{\pm}/Z$ data in the whole $x$ region.
 In the small-$x$ region $10^{-5}<x<10^{-2}$ especially, the uncertainty would be reduced by a factor of 60\% at $x\sim10^{-3}$.
 \item The uncertainty of $u$, $s$, $c$ quark and gluon PDFs can be reduced across the whole $x$ region, 
 and significant improvements are expected in very small- and larger-$x$ regions.
 \item The uncertainties of \doveru and $\bar{d}/\bar{u}$ ratios can be significantly 
 reduced across the whole $x$ range, even with only 5\invfb data. In the very larger-$x$ region, 
 the LHCb 13\tev data could have a large impact on the \doveru ratio.
 \item The LHCb 13\tev $W^{\pm}$ and $Z$ data also has a large impact on the $\bar{u}$ and $\bar{d}$ quark PDFs, mainly in the small-$x$ region.
 \end{itemize}

For the \doveru and $\bar{d}/\bar{u}$ ratios, the future LHCb 13\tev data will provide the most important constraints on them. 
In Fig.~\ref{fig:wz_inputs_udratio}, the 300\invfb LHCb 13\tev pseudo-data provides valuable constraints on the \doveru ratio in the very 
large-$x$ region ($> 0.5$) and $\bar{d}/\bar{u}$ in $x > 0.2$. 
In these regions, the LHCb data would be the only clean data, as it is free of nuclear corrections, 
as needed when describing the low energy Drell-Yan data to constrain \doveru. 

\begin{figure}[!htbp]
\centering
\includegraphics[width=0.65\textwidth]{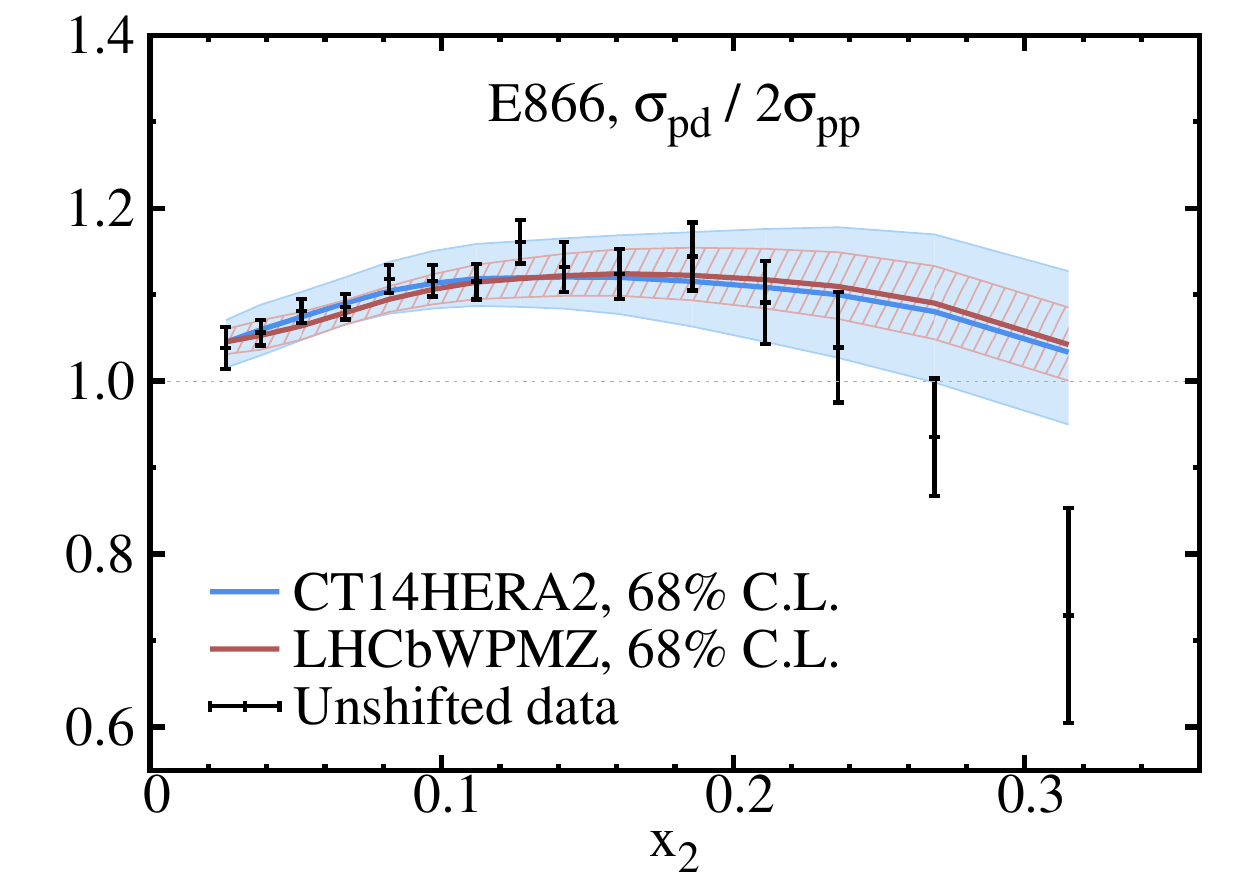}
\caption{Theoretical predictions based on the updated CT14HERA2 (red band) and original CT14HERA2 (blue band)
for the fixed-target Drell-Yan cross section, $\sigma_{pd} \big/ 2\sigma_{pp}$, in the region of larger $x_2 \gtrsim 0.1$
to be probed by the SeaQuest experiment~\cite{Aidala:2017ofy} at Fermilab.
For comparison, the higher-$x_2$ portion of the older E866 data~\cite{Towell:2001nh} (black points) is also presented here.}
\label{fig:hix_DY}
\end{figure}

It is well-known that fixed-target Drell-Yan measurements provide an important probe of the $x$ dependence of the nucleon
(and nuclear) PDFs. This fact has motivated a number of experiments, including the Fermilab E866/NuSea
experiment \cite{Towell:2001nh}, which determined the normalized deuteron-to-proton cross section ratio
$\sigma_{pd} \big/ 2\sigma_{pp}$ out to relatively large $x_2$, the momentum fraction of the target. As can be seen
based upon a leading-order quark-parton model analysis, this ratio is expected to have
especially pronounced sensitivity to the $x$ dependence of the PDF ratio $\bar{d}/\bar{u}$.
The E866 results stimulated an interest in performing a similar measurement out to larger $x_2$ with higher
precision --- the main objective of the subsequent SeaQuest/E906 experiment at Fermilab \cite{Aidala:2017ofy}, from which
results are expected soon.
The LHCb data could be used to check the impact from 
the SeaQuest~\cite{Tadepalli:2019ghd} result for $\bar{d}/\bar{u}$ in the large-$x$ region. 
In Fig.~\ref{fig:hix_DY}, we compare the theoretical prediction based on the update CT14HERA2 PDFs 
(with the 300\invfb LHCb 13\tev combined $W^\pm$ and $Z$ pseudo-data as input) 
and the original CT14HERA2 PDFs. This shows that the LHCb 13\tev data could further constrain the deuteron-to-proton ratio, 
$\sigma_{pd} \big/ 2\sigma_{pp}$, in the large-$x$ region.

%%%%%%%%%%%%%%%%%%%%%%%%%%%%%%%%%%%%%%%%%%%%%%%%
\subsection{Update from LHCb 13 TeV $(W^++W^-)/Z$ pseudo-data } \label{subsec:s-PDF}
	
At tree level, the $Z$ boson is produced via
$q\bar{q}$ annihilation, where $q$ could be $u$, $d$, $c$, $s$, and $b$,
while the $W^\pm$ boson could produced via $u\bar{s}$ and $\bar{u}s$.
Therefore, the ratio of $W^\pm$ distribution to that of the $Z$ boson could be sensitive to the \ssoverud at first order~\cite{Nadolsky:2008zw}.

\begin{figure}[!htbp]
\centering
\includegraphics[width=0.45\textwidth]{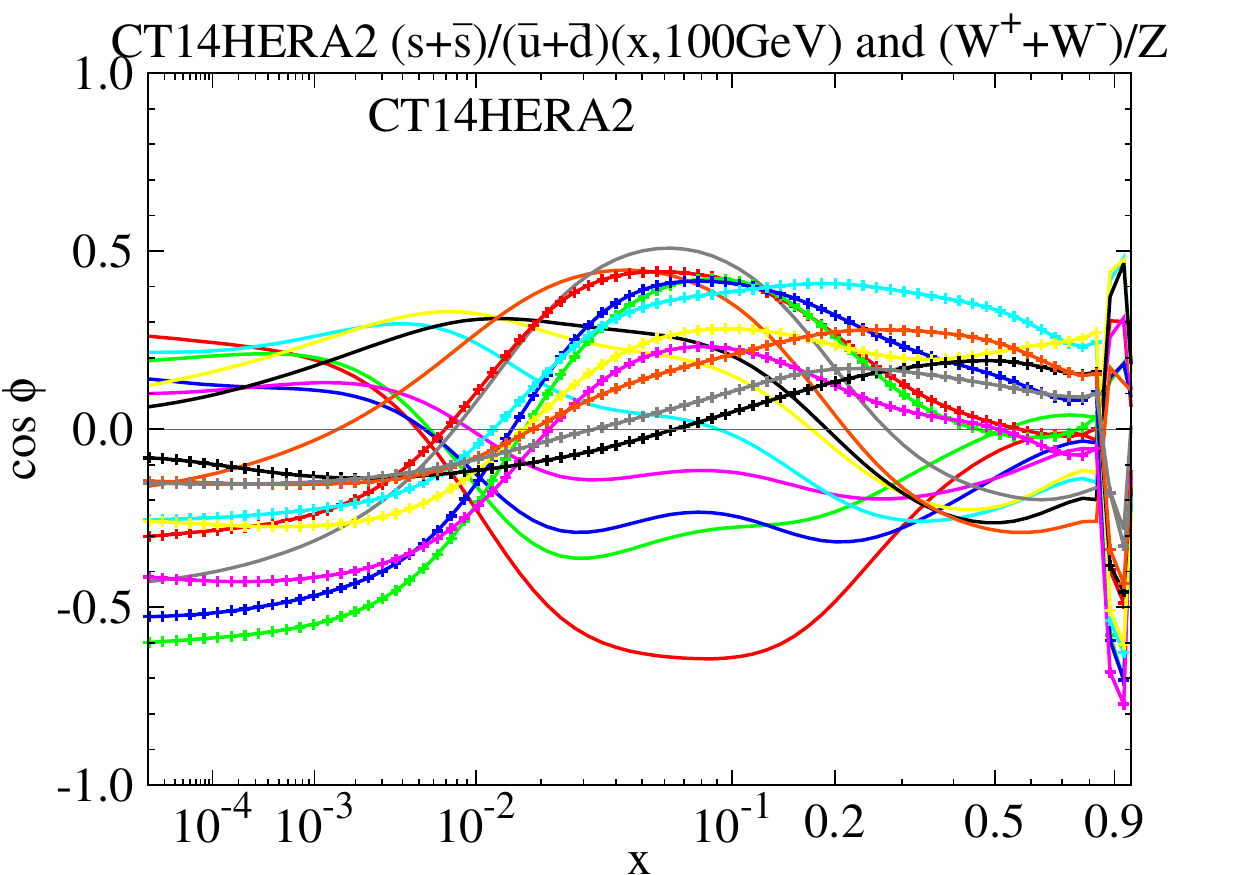}
\includegraphics[width=0.45\textwidth]{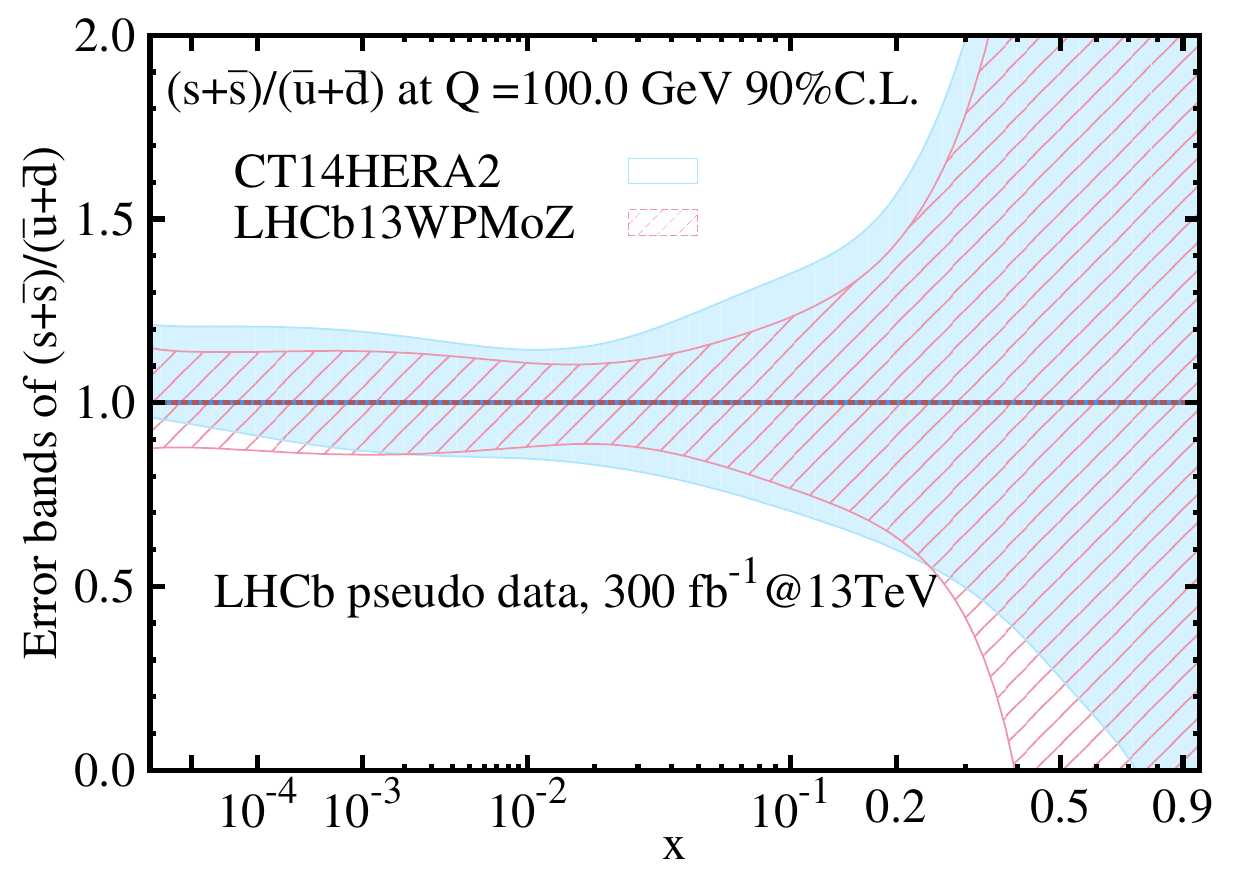}
\caption{
(left) $\cos\phi$ between the PDF ratio \ssoverud and the $(W^++W^-)/Z$ ratio. 
Lines with different color represent different bins, and the same index as Fig.~\ref{fig:z_x_corr} is used.
(right) The PDF uncertainties associated with \ssoverud distribution, using the 300\invfb data, as a function of $x$, 
in the CT14HERA2 PDFs and {\sc ePump} updated new PDFs. 
The denominator is the central value of each PDF set.
The single $(W^++W^-)/Z$ boson differential cross sections are taken as the {\sc ePump} input.}
\label{fig:ssbar}
\end{figure}

With a uniform binning (18 pseudorapidity/rapidity bins, from 2.0 to 4.5), a $(W^++W^-)/Z$ ratio in each bin is calculated, 
where the muon pseudorapidity of the $W^{\pm}$ boson and the rapidity of the $Z$ boson are used. 
Correlations between the predicted LHCb 13\tev $(W^++W^-)/Z$ ratio and \ssoverud are shown in Fig.~\ref{fig:ssbar}.
With the calculated $(W^++W^-)/Z$ ratio as {\sc ePump} input, we checked the impact on \ssoverud from the LHCb single $W^{\pm}$ and $Z$ data, as shown in Fig.~\ref{fig:ssbar}.
With 5\invfb LHCb 13\tev pseudo-data as input, the $(W^++W^-)/Z$ data does not have a visible impact on the PDF ratio \ssoverud.
With 300\invfb LHCb 13\tev pseudo-data as input, the impact on the PDF ratio \ssoverud
becomes significant in the $x$ range of $10^{-2}$ to $10^{-1}$, 
which could be used to precisely determine the strange quark PDFs in the future. 
As expected, larger correlations between $(W^++W^-)/Z$ and \ssoverud are seen in the same $x$ range.

%%%%%%%%%%%%%%%%%%%%%%%%%%%%%%%%%%%%%%%%%%%%%%%%
\section{Tolerance criteria in {\sc ePump} update}\label{sec:tolerance}

The tolerance criterion ({\it i.e.}, the choice of total $\Delta \chi^2$ value in a global analysis) is an important parameter in PDF fitting.

It was extensively discussed in Ref.~\cite{Hou:2019efy} that in order to best reproduce CT14HERA2 global fit, one should use dynamical tolerance in {\sc ePump}. 
If the tolerance is set to be $\Delta \chi^2=1$ at 68\% CL, or equivalently, $(1.645)^2$ at 90\% CL,
instead of using a dynamical tolerance, a very large weight is effectively assigned to the new input data, when updating the existing PDFs in the CT PDF global analysis framework.

\begin{figure}[!htbp]
\centering
\includegraphics[width=0.45\textwidth]{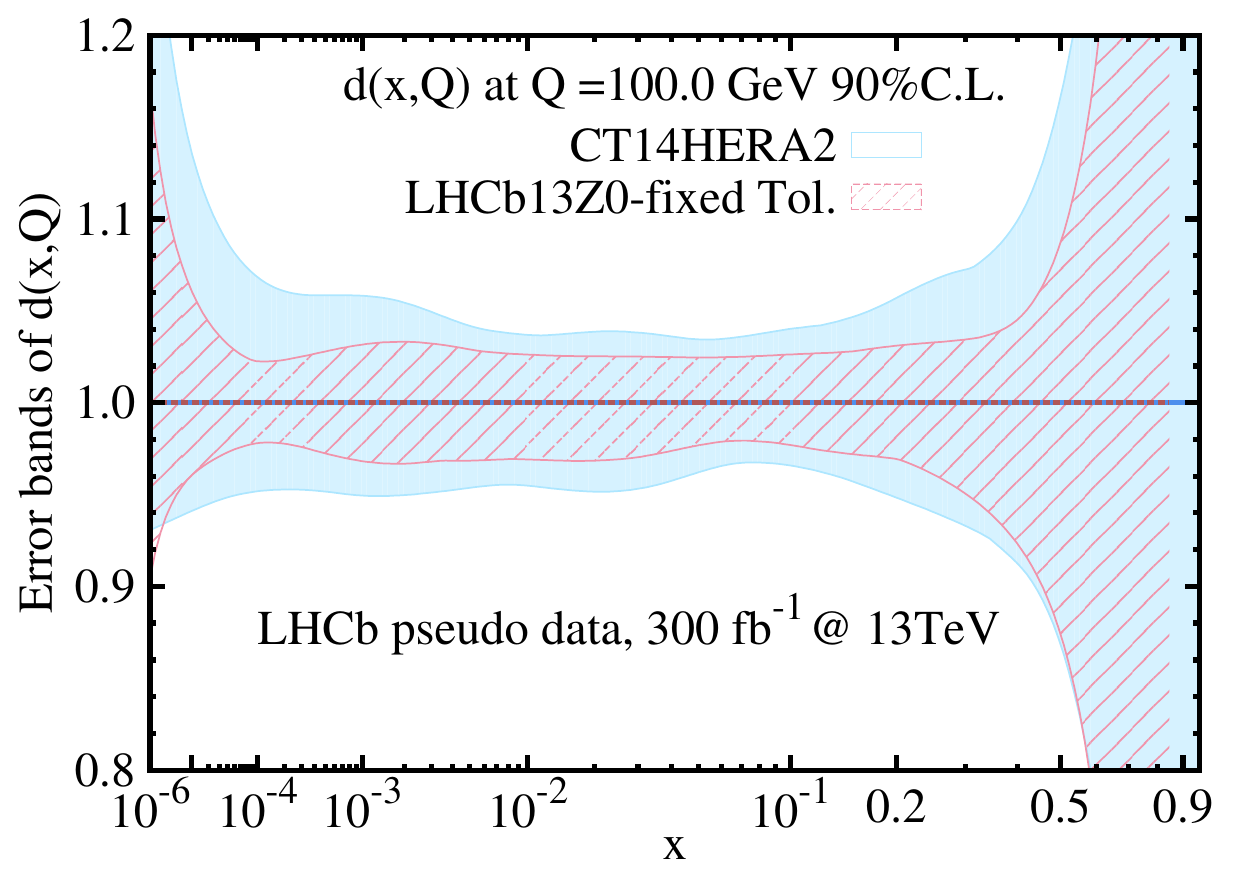}
\includegraphics[width=0.45\textwidth]{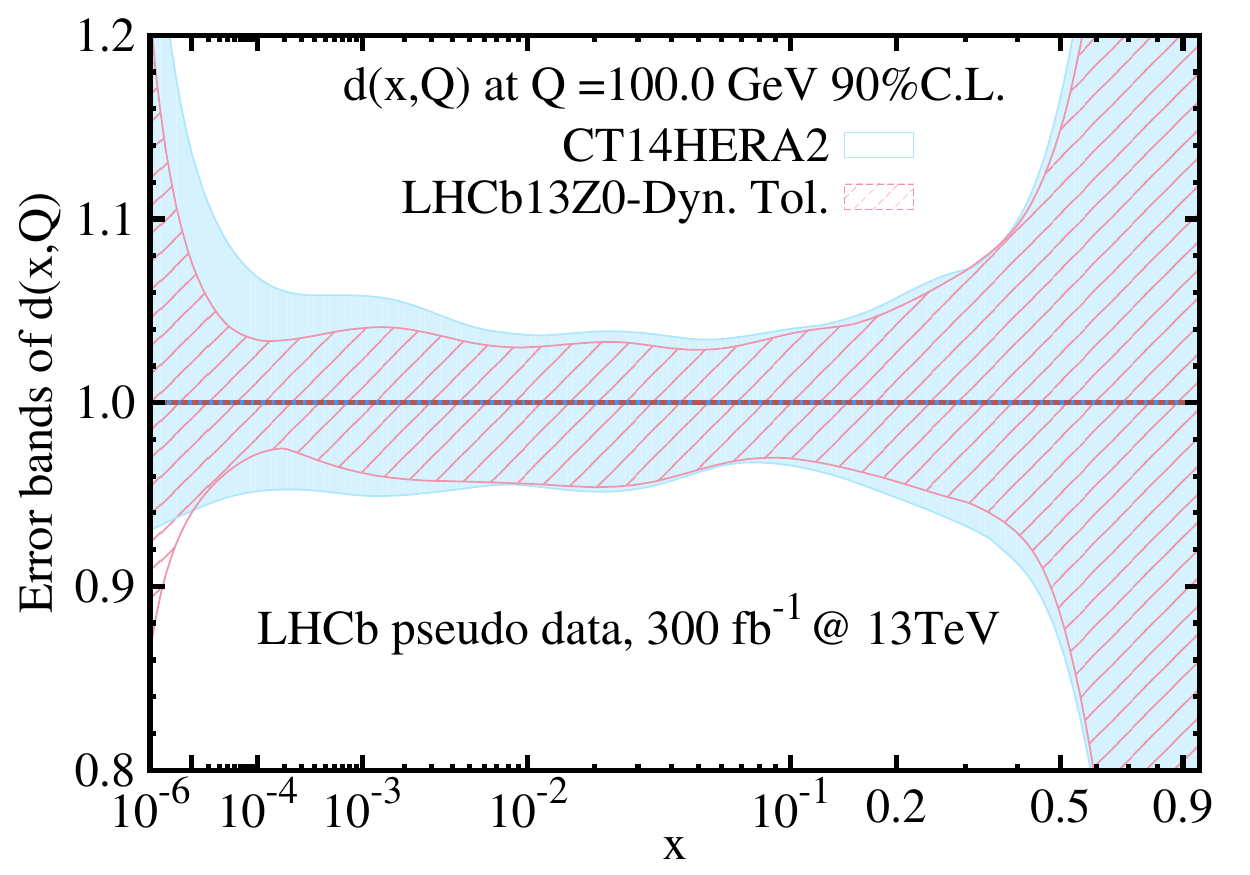}
\caption{PDF uncertainties associated with the $d$ quark distribution, as a function of $x$, in the CT14HERA2 PDFs and {\sc ePump} updated new PDFs. 
The denominator is the central set of each PDF sets.
The rapidity distribution of $Z$ boson events (300\invfb) is used as input to the {\sc ePump} update. (left) The result with fixed tolerance of $\Delta\chi^2=(1.645)^2$ at the 90\% CL, and (right) the result with dynamical tolerance.
\label{fig:z0_udquark_dyn_1}}
\end{figure}

To illustrate the differences between the {\sc ePump} updated PDFs using a dynamical 
tolerance and a fixed tolerance with $\Delta \chi^2=(1.645)^2$, we used the single LHCb 13\tev $Z$ data as the {\sc ePump} input.
The result is shown in Fig.~\ref{fig:z0_udquark_dyn_1}. 
The impact from the new data is enhanced when the PDFs are updated with $\Delta \chi^2=(1.645)^2$, which could introduce bias in the new updated PDF set. 

This conclusion also holds for using MMHT2014~\cite{Harland-Lang:2014zoa} and PDF4LHC15~\cite{Butterworth:2015oua} PDFs in profiling analysis to study the impact of a new (pseudo-) data on updating the existing PDFs. 

%%%%%%%%%%%%%%%%%%%%%%%%%%%%%%%%%%%%%%%%%%%%%%%%
\section{Conclusion}\label{sec:summary}

With detector instrumented in the forward region, the $pp$ collision data collected by the LHCb experiment provides essential and 
complementary information for a global analysis of experimental data to determine the PDFs of the proton, as 
compared to data collected by the ATLAS and CMS detectors.

In this article, we have studied the potential of the LHCb 13\tev single \wz boson pseudo-data 
for constraining the PDFs in the proton. 
As an example, we demonstrated the sensitivity of the LHCb 13\tev data, collected with an integrated luminosity of 5\invfb and 300\invfb, respectively, to reducing the PDF uncertainty bands of the CT14HERA2 PDFs. We have also investigated the sensitivities of various experimental observables. 

A large impact from the LHCb 13\tev data on various quark flavor PDFs has been seen across the $x$ region, 
and significant contributions in the small-$x$ ($<10^{-3}$) region are expected. In particular, the $d$ and $\bar{d}$ quark PDF uncertainties are reduced dramatically, 
$\sim$60\% improvement at momentum fraction ($x$) around $10^{-3}$.
Due to its large event rate, the LHCb 13\tev ${W^+}/{W^-}$ data can already further reduce the \doveru PDF uncertainty, even with only a 5\invfb data sample, as shown in Fig.~\ref{fig:udratio_wpm}. 
Most of the improvements are concentrated in the small-$x$ region, from $10^{-5}$ to $10^{-3}$. 
The 300\invfb LHCb 13\tev data sample could further reduce the uncertainties of both the PDF ratios \doveru and \doverubar (by $\sim 20\%$)
in the small-$x$ region, as well as giving some noticeable improvements in the large-$x$ region, which is currently dominated by DIS data in global analysis. 

Although it has a smaller event rate than the $W^\pm$ events, the LHCb 13\tev $Z$ data can also provide important constraints on PDFs, particularly when considering a double-differential distribution (as a function of $Z$ boson \pt and rapidity), as the experimental observable for updating the existing PDFs. 
We have compared the impact of the LHCb 13\tev $Z$ boson pseudo-data on PDFs by performing 
single differential ($Z$ boson \pt), 
double-differential ($Z$ boson \pt and $Z$ boson rapidity),
and triple-differential ($Z$ boson \pt, $Z$ boson rapidity, and lepton $\cos\theta^*$ ) 
cross section measurements.
We found that 
with limited statistics of $Z$ boson events (5\invfb data), 
the mulit-dimensional measurements cannot significantly improve the PDF determination, as compared to one-dimensional measurements of 
$Z$ boson \pt, lepton $\cos\theta^*$, and $Z$ boson rapidity, respectively. 
With a 300\invfb data sample, however, the mulit-dimensional measurement has 
better constraints on the PDFs, across the whole $x$ range.

It is evident that the combined data sample of $W^\pm$ and $Z$ boson events can provide further constraints on PDFs. Examining the ratio of the event rates 
$(W^++W^-)/Z$ could directly probe the PDF ratio \ssoverud, as previously noted in Ref.~\cite{Nadolsky:2008zw}.
With 5\invfb LHCb 13\tev pseudo-data as input, the $(W^++W^-)/Z$ data does not have a visible impact on the PDF ratio \ssoverud.
With 300\invfb LHCb 13\tev pseudo-data as input, the impact on the PDF ratio \ssoverud
becomes significant in the $x$ range of $10^{-2}$ to $10^{-1}$, which could be used to precisely determine the strange quark PDFs in the future.
The above features suggest that the LHCb single $W^{\pm}$ and $Z$ data taken at the LHC 13\tev will provide very important and unique information in a global analysis, complementary to the ATLAS and CMS results.
With an integrated luminosity of 300\invfb data to be collected in the future, the impact from the LHCb 13\tev data on the PDF fitting could be enhanced significantly by performing a multi-differential cross section measurement. 

Before concluding our work, we also pointed out the important role of tolerance criteria in the PDF updating.
As discussed in Ref.~\cite{Hou:2019efy}, one should use dynamical tolerance in {\sc ePump}. 
Setting the tolerance to be 1 at 68\% CL, or equivalently, $(1.645)^2$ at 90\% CL, will greatly overestimate the impact of a given new data set when updating the existing PDFs in the CT PDF global analysis framework.
This conclusion also holds for using MMHT2014~\cite{Harland-Lang:2014zoa} and PDF4LHC15~\cite{Butterworth:2015oua} PDFs in profiling analysis to study the impact of a new (pseudo-) data on updating the existing PDFs. 

%%%%%%%%%%%%%%%%%%%%%%%%%%%%%%%%%%%%%%%%%%%%%%%%
\section{Acknowledgments}
This work is supported by the National Natural Science Foundation of China under the Grant No. 11875142, No. 11965020, 
and by the U.S.~National Science Foundation under Grant No.~PHY-2013791. 
C.-P.~Yuan is also grateful for the support from the Wu-Ki Tung endowed chair in particle physics.

%%%%%%%%%%%%%%%%%%%%%%%%%%%%%%%%%%%%%%%%%%%%%%%%

%%%%%%%%%%%%%%%%%%%%%%%%%%%%%%%%%%%%%%%%%%%%%%%%
\end{document}